\documentstyle[epsf,aps,preprint,tighten]{revtex}

\begin{document} 

\title{A nonperturbative determination of the $O(a)$ improvement
coefficient $c_A$ and the scaling of $f_\pi$ and
$m^{\overline{MS}}$.}

\author{UKQCD Collaboration}
\author{S.~Collins and C.~T.~H.~Davies}
\address{University of Glasgow, Glasgow, G12 8QQ,
Scotland, UK}
\author{G.~P.~Lepage}
\address{Newman Laboratory of Nuclear Studies, Cornell University,
Ithaca, NY 14853 USA}
\author{J.~Shigemitsu}
\address{The Ohio-State University, Columbus, OH 43210, USA}

\maketitle
\narrowtext
\begin{abstract}
We report on an investigation of the LANL method for determining the
$O(a)$ improvement coefficient $c_A$ nonperturbatively. We find we
are able to extract reliable estimates for the coefficient using this
method. However, our study of systematic errors shows that for very
accurate determinations of $c_A$, the smearing function must be tuned
and the volume fixed to keep the $O(a)$ ambiguity in $c_A$ fixed as
$\beta$ varies. Consistency was found with previous results from the
LANL group and (within fairly large errors) 1-loop perturbation
theory; $c_A$ does not change significantly over the range
$\beta=5.93{-}6.2$. The big difference between our results and those
of the ALPHA collaboration, around $\beta=6.0$, show that the $O(a)$
differences in $c_A$ between the different methods can be large. We
find that the lattice spacing dependence of $f_\pi$ and the
renormalised quark mass is much smaller using our values of the
coefficient compared to those of the ALPHA collaboration.
\end{abstract}

\section{Introduction}
The use of Symanzik improvement~\cite{syman} of lattice actions and
matrix elements is widespread and very effective. However, with each
improvement term added the corresponding coefficient must be
determined to enable discretisation effects to be reduced to the
desired level. Considering the light hadron spectrum and matrix
elements, the relevant $O(a)$ improvement coefficients are, for the
most part, only known to $1$-loop in perturbation theory, leaving
residual $O(\alpha^2 a)$ discretisation terms. A nonperturbative
determination of these coefficients is desirable to completely remove
$O(a)$ effects. Such a determination is possible through the
imposition of the axial Ward identities~(AWI) on the lattice.

Central to a programme of determining $O(a)$ improvement coefficients
nonperturbatively is the calculation of $c_A$, the improvement
coefficient of the axial-vector current; the improved current appears
in the expression for the generic axial Ward identity and $c_A$ must
be determined before other operator improvement coefficients can be
calculated~\cite{lanl2}. So far, two groups, the LANL
group~\cite{lanl,lanl2,rajan} and the ALPHA
collaboration~\cite{alpha,bpba}, have calculated $c_A$ along with
several other improvement coefficients.

\begin{figure}
\begin{center}
\epsfxsize=8.0truecm\epsffile{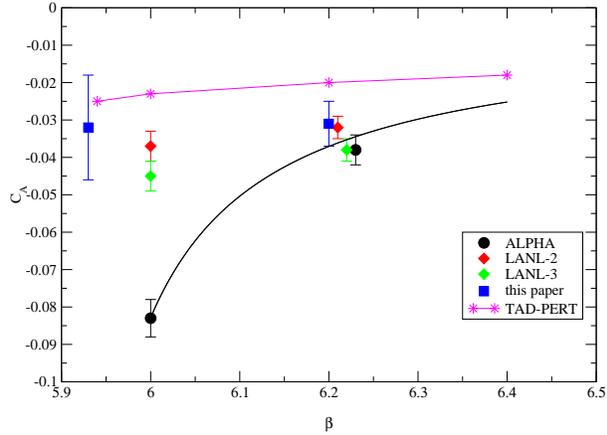}
\caption{$C_A$ extracted by various groups. The 1-loop  tadpole-improved values for $c_A$ were calculated using $\alpha_P(1/a)$. Also shown as a black line is the Pade expansion of the ALPHA results~\protect\cite{alpha}. The ALPHA and LANL results at $\beta=6.2$ are offset for clarity. LANL-3 and LANL-2 refer to the results of the LANL group using standard $O(a^2)$ derivatives~(equation~\protect\ref{norm}) and modified $O(a^2)$ derivatives~\protect\cite{lanl} respectively.}\label{falpha}
\end{center}
\end{figure}

Their results for $c_A$ at $\beta=6.0$ and $6.2$ are summarised in
figure~\ref{falpha} and compared with 1-loop perturbation
theory~(using $\alpha=\alpha_P(1/a)$~\cite{alphap}). While the results
are compatible on the finer lattice, there is a big difference in the
values for $\beta=6.0$. This difference can be explained by the $O(a)$
ambiguity which exists in nonperturbative determinations of $O(a)$
improvement coefficients. As long as the improvement conditions for
each determination are applied consistently as $\beta$ changes,
differences in the value of $c_A$ are not important in principle; the
difference will disappear in the continuum limit.

In practice, different values of $c_A$ can have a large effect away
from the continuum limit. This is because the matrix elements appearing
at $O(a)$ for $f_\pi$ and the renormalised quark mass are numerically
large compared to the leading order term.
\begin{eqnarray}
f^{imp} & = & \left<P|A_4|0\right>+ c_A\left<P|a\partial_4P|0\right>\label{fimp}\\
m_{PCAC}^{imp} & = & \frac{\left<P|\partial_4A_4|0\right>}{2\left<P|P|0\right>} + c_A \frac{\left<P|a\partial_4^2P|0\right>}{2\left<P|P|0\right>}, \label{mpcac}
\end{eqnarray}
and $a\left<P|\partial_4P|0\right>/\left<P|A_4|0\right>\sim a
M_\pi^2/m_q$.  An $O(a)$ ambiguity in $c_A$ therefore appears at
$O(a^2)$ but multiplied by a large matrix element and it is undesirable
to have large $O(a^2)$ scaling violations, even if $O(a)$ errors
have been removed.

Our aim is to investigate how well $c_A$ is determined using the LANL
method. The latter only requires a conventional analysis, which is
available from simulations performed for spectroscopic calculations,
compared to using the Schr\"odinger functional techniques of the ALPHA
collaboration. With significantly higher statistics than those of
reference~\cite{lanl2} we are able to improve on the LANL analysis by
performing correlated fits, investigating the choice of lattice
derivatives employed more widely and determining the stability of the
LANL results to changes in the fitting range. In addition, we study
the scaling behaviour of $f_\pi$ and the renormalised quark mass with
respect to the choice made for $c_A$.

The paper is organised as follows: in section~\ref{capcac} we outline
how to extract $c_A$ from the PCAC relation and in particular the
method employed by the LANL group. Results are presented in
section~\ref{caresults}, which includes a comparison of our results
with those of the LANL group and the ALPHA collaboration. The scaling
of $f_\pi$ and the renormalised quark mass is dealt with in
section~\ref{scalingsec}, followed by the conclusions in
section~\ref{conc}. Technical details not directly related to the
method of calculating $c_A$ - the simulation details, extracting meson
masses and decay constants, the renormalisation factors used to obtain
$f_\pi$ and $m^{\overline{MS}}$ and the chiral extrapolations are all
given in the Appendix.

\section{$c_A$ from the PCAC relation}
\label{capcac}
\begin{figure}
\begin{center}
\epsfxsize=8.0truecm\epsffile{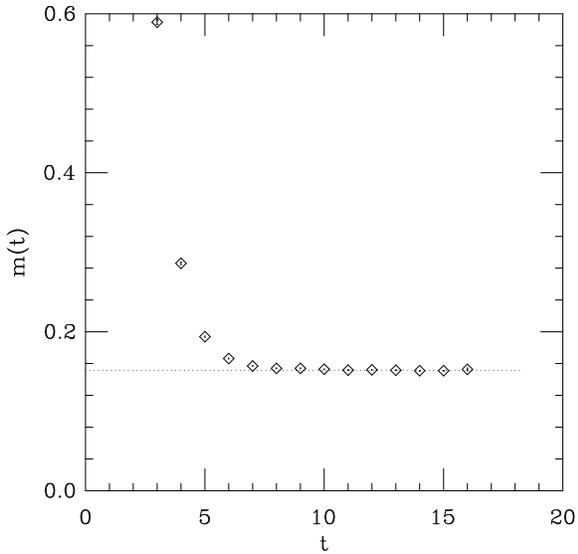}
\caption{$m(t)$ as a function of timeslice for $\beta=5.93$,
 $\kappa_l=0.1327$ and $LL$ smearing. The dotted line indicates a fit
 to a constant.}\label{mfig}
\end{center}
\end{figure}

The PCAC relation in euclidean space can be written as
\begin{equation}
\left<J\partial_\mu A_\mu(x) \right> = 2m_{PCAC} \left<JP(x)\right>\label{pcac}
\end{equation}
and should hold on the lattice for all $x$ not coincident with $J$ up to
discretisation terms.  The axial-vector current, $A_\mu$, is given by
$\bar{\psi}\gamma_\mu\gamma_5\psi$, the pseudoscalar operator
$P=\bar{\psi}\gamma_5\psi$ and $J$ is any operator with the
pseudoscalar quantum numbers. $m_{PCAC}$ is the bare current quark
mass. For simplicity we sum over position space, restricting ourselves
to zero momentum, and define
\begin{eqnarray}
r_J(t) & = & \frac{<J\partial_4 A_4(t) >}{<JP(t)>}\label{er}.
\end{eqnarray}
Thus, equation~\ref{pcac} becomes
\begin{equation}
 r_J(t) = m(t) = 2m_{PCAC}+O(a).
\end{equation}
This relation holds for all states~(ground and radial excitations) of
the pseudoscalar meson.  In the limit of large times, when only the
ground state contributes to $r_J$, then $m(t)$ is a constant given by
$2m_{PCAC}+O_{g.s.}(a)$, where $O_{g.s.}(a)$ are the discretisation
errors associated with the ground state. At earlier times, when
excited states make a significant contribution~(with different
discretisation errors), $m(t)$ becomes time dependent, as can be seen
in figure~\ref{mfig}.

The size of the discretisation terms~(and the time dependence of
$m(t)$) are reduced to $O(a^2)$ when we improve the axial-vector
current~\footnote{The quark action must also be improved to $O(a^2)$
using the Sheikholeslami-Wohlert term with the value of $c_{SW}$
determined nonperturbatively.}
\begin{equation}
A_4\rightarrow A_4^I = A_4 + ac_A\partial_4 P + O(a^2)\label{improv}.
\end{equation}
Then
\begin{equation}
r_J(t) + ac_As_J(t) = m_{imp}(t) = 2m_{PCAC} + O(a^2)\label{pcacimp},
\end{equation}
with
\begin{eqnarray}
s_J(t) & = & \frac{<J\partial_4^2 P(t) >}{<JP(t)>}\label{es}.
\end{eqnarray}
Clearly, changing $J$ or the time $t$ changes the size of the
contribution of each state to $r_J$, and the size of $O(a)$ terms. The
improvement term must still cancel these terms, however, giving rise
to a quark mass which differs from $m_{imp}$ only in $O(a^2)$.
\begin{equation}
r_J(t') + ac_As_J(t') = m_{imp}(t') = 2m_{PCAC} + O'(a^2),
\end{equation}
or
\begin{equation}
r_{J'}(t) + ac_As_{J'}(t) = m_{imp}'(t) = 2m_{PCAC} + O''(a^2). 
\end{equation}
By forcing $m_{imp}(t)$ and $m_{imp}(t')$ to be equal we
can solve for $c_A$.
\begin{eqnarray}
 -\frac{1}{a} \frac{r(t') - r(t)}{s(t')-s(t)} & \equiv & c_A .\label{calanl}
\end{eqnarray}
A suitable choice for $t$ and $t'$ is to set $t=t_{gs}$ in the region
where the ground state dominates $r_J$ and $s_J$ and $t'=t_{ex}$ in
the region where there is significant contribution from excited
states.

In order to illuminate the $O(a)$ ambiguity in $c_A$, we identify
\begin{eqnarray}
\Delta r_J & = & r_J(t') - r_J(t)= \Delta r_J[O(a)] + \Delta r_J[O(a^2)] +\ldots\\
\Delta s_J & = & s_J(t')-s_J(t) \approx  -\Delta r_J[O(a)]/ac_A\\
 -\frac{1}{a} \frac{r_J(t') - r_J(t)}{s_J(t')-s_J(t)} & \approx & c_A + c_A \frac{\Delta r_J[O(a^2)]}{\Delta r_J[O(a)]}
\end{eqnarray}
where $\Delta r_J[O(a)]$ denotes the change in $r_J$ due to $O(a)$
effects etc.  Thus, the $O(a)$ ambiguity in $c_A$ depends on the
difference of the $O(a^2)$~(and $O(a)$) terms between $r$ at $t_{ex}$
and $t_{gs}$ rather than the absolute values. Obviously if $\Delta
r_J[O(a^2)]\sim\Delta r_J[O(a)]$, the error in $c_A$ will be as large as
$c_A$ itself~(and the above expansion will not be valid).

The LANL method is equivalent to using equation~\ref{calanl}. It
involves performing a fit to $r_J(t)$ and $s_J(t)$ such
that $r_J(t) + ac_As_J(t)$ is equal to a constant~($2m$), where $c_A$
and $2m$ are parameters in the fit. The fitting range is chosen to be
from $t_{ex}$ to $t_{gs}$. The advantage of performing a fit over
calculating the ratio in equation~\ref{calanl} is that one can test
the ansatz that the value of $c_A$ reduces the time dependence of
$r_J(t)$~(and hence the discretisation errors) with the
$\chi^2$.

We note that the ALPHA collaboration employs a slightly different
method to calculate $c_A$. Within the Schr\"odinger functional
approach it is possible to simply work in the region when the ground
state dominates and $r_J$ and $s_J$ have plateaued. The boundary
fields are varied and this changes the discretisation errors in the
ground state. A ratio similar to equation~\ref{calanl} is built up
from $r_J$ and $s_J$ from different boundary fields. A feature of
using the Schr\"odinger functional technique is that the analysis can
be performed at directly zero quark mass.

The unknown $O(a)$ ambiguity in $c_A$ is due to $O(a^2)$ errors in the
axial-vector current, the pseudoscalar current and the light quark
action. The improvement scheme which we have chosen defines $c_A$ in
the limit of zero quark mass, however, $c_A$ is calculated at finite
$m_q$ and then extrapolated. There is an additional $O(a)$ ambiguity due
to the discretisation chosen for the temporal derivatives in
equations~\ref{er} and~\ref{es}, which is proportional to $m_q$ and
vanishes in the chiral limit. As part of our analysis we investigated
two different choices for the lattice derivatives in the determination
of $c_A$: ``standard'' symmetric lattice derivatives
\begin{eqnarray}
\partial_{\mu} \rightarrow \Delta_{\mu}^{(+-)} & = & \frac{1}{2}(\delta_{\vec{x},\vec{x}+\hat{\mu}}-\delta_{\vec{x},\vec{x}-\hat{\mu}})\label{norm}\\
\partial^2_{\mu} \rightarrow \Delta^{(2)}_{\mu} & = & \delta_{\vec{x},\vec{x}+\hat{\mu}} - 2\delta_{\vec{x},\vec{x}} + \delta_{\vec{x},\vec{x}-\hat{\mu}}\nonumber
\end{eqnarray}
which contain $O(a^2)$ errors, and ``improved'' $O(a^4)$ derivatives
\begin{eqnarray}
\tilde{\Delta}_{\mu}^{(+-)} & = & \Delta_{\mu}^{(+-)} -\frac{1}{6}\Delta^+\Delta^{(+-)}\Delta^{-}\label{impd}\\
\tilde{\Delta}^{(2)}_{\mu} & =& \Delta_{\mu}^{(2)} -\frac{1}{12}\left[\Delta^+\Delta^-\right]^2\nonumber
\end{eqnarray}
where $\Delta^{+} = \delta_{\vec{x},\vec{x}+\hat{\mu}} - 1$ and
$\Delta^{-} = 1- \delta_{\vec{x},\vec{x}-\hat{\mu}}$.  Two points
should be taken into account when choosing the form for the
derivatives. As one improves the derivatives then the minimum value
that $t_{ex}$ can take becomes larger since there must be no overlap
with the origin.  In addition, noise rapidly dominates the
determination of $c_A$ as $t_{ex}$ increases and there is only a small
window of timeslices from which to extract $c_A$. The LANL group
considered another form of $O(a^2)$ lattice derivative~\cite{lanl}
which has smaller $O(a^2)$ terms compared to eqns~\ref{norm} but uses
fewer timeslices than the $O(a^4)$ derivatives. While the different
definitions give consistent results for $c_A$ in the chiral limit, we
found using improved derivatives helped in extracting $c_A$.  This
point is discussed in the next section.

In addition to the LANL method and equation~\ref{calanl} we considered
extracting $c_A$ by changing $J$ from $\bar{\psi}\gamma_5\psi$ to
$\bar{\psi}\gamma_4\gamma_5\psi$. The change in $J$ does not give rise
to significantly different discretisation effects and the
determination of $c_A$ was not improved. In the following we set $J=P$
and drop the subscript on $r$ and $s$. We also used the LANL method
with finite momentum correlators in $r$ and $s$~(including the
additional spatial derivative terms). However, at finite momentum the
$O(a)$ errors in $c_A$ are increased and no additional constraint on
the coefficient is obtained compared to the zero momentum results.

\section{Results for $c_A$}
\label{caresults}
We have tested the LANL method using the UKQCD quenched data set. The
simulation details are given in table~\ref{sim} and discussed in the
Appendix. The best analysis was possible at $\beta=5.93$.

\subsection{Results at $\beta=5.93$}
\begin{figure}
\begin{center}
\centerline{
\epsfxsize=8.0truecm\epsffile{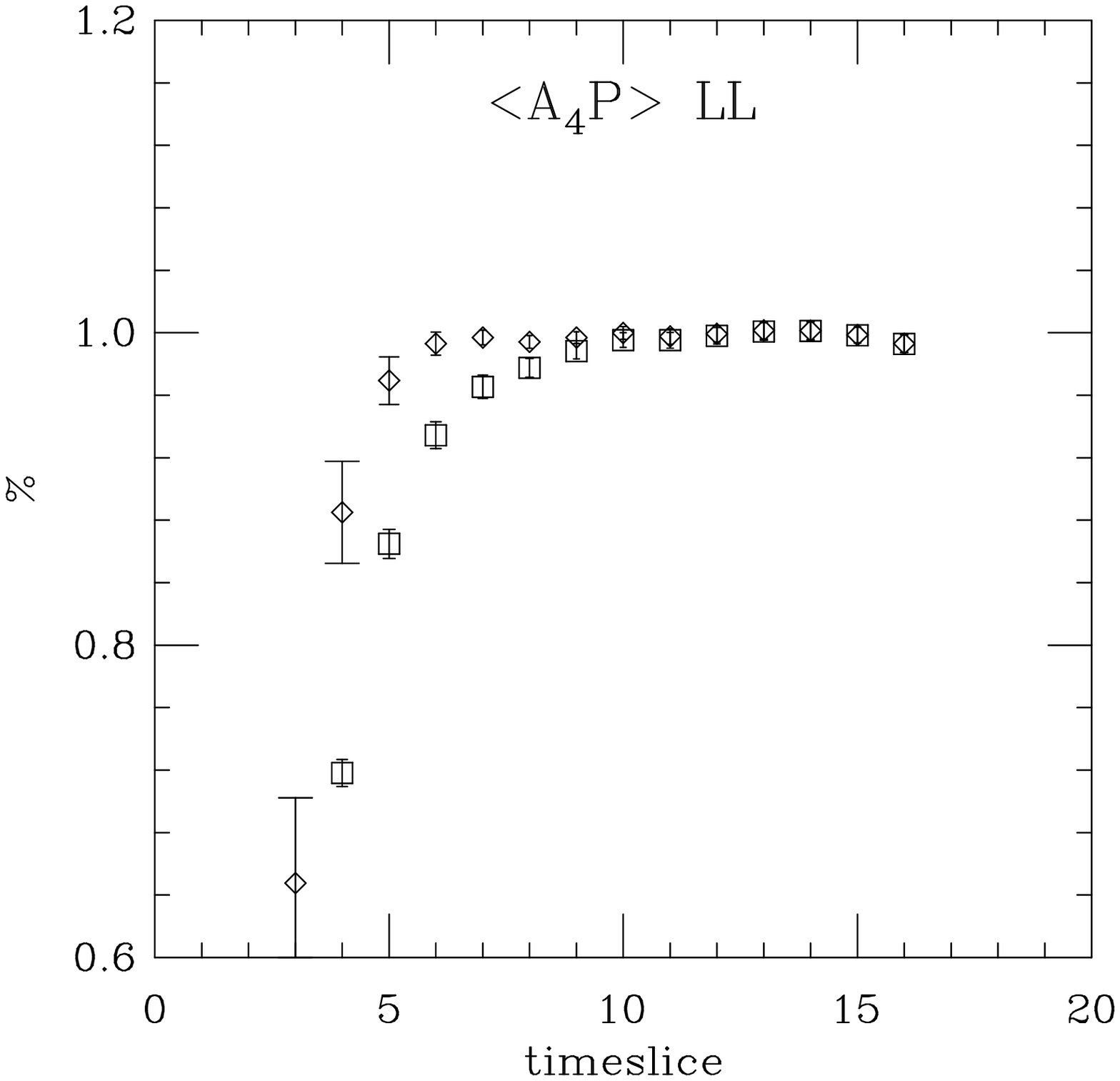}
\epsfxsize=8.0truecm\epsffile{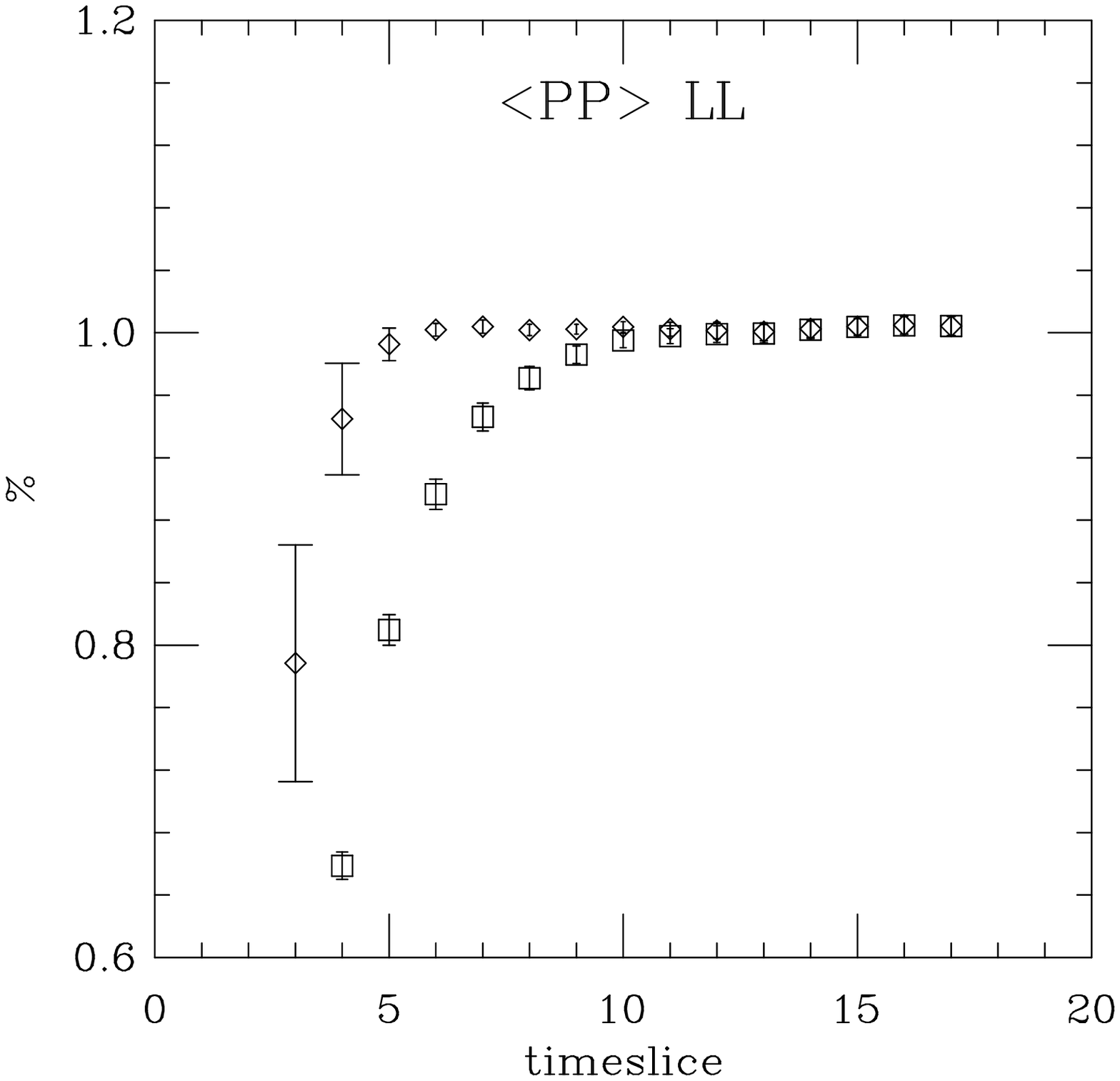}}
\centerline{
\epsfxsize=8.0truecm\epsffile{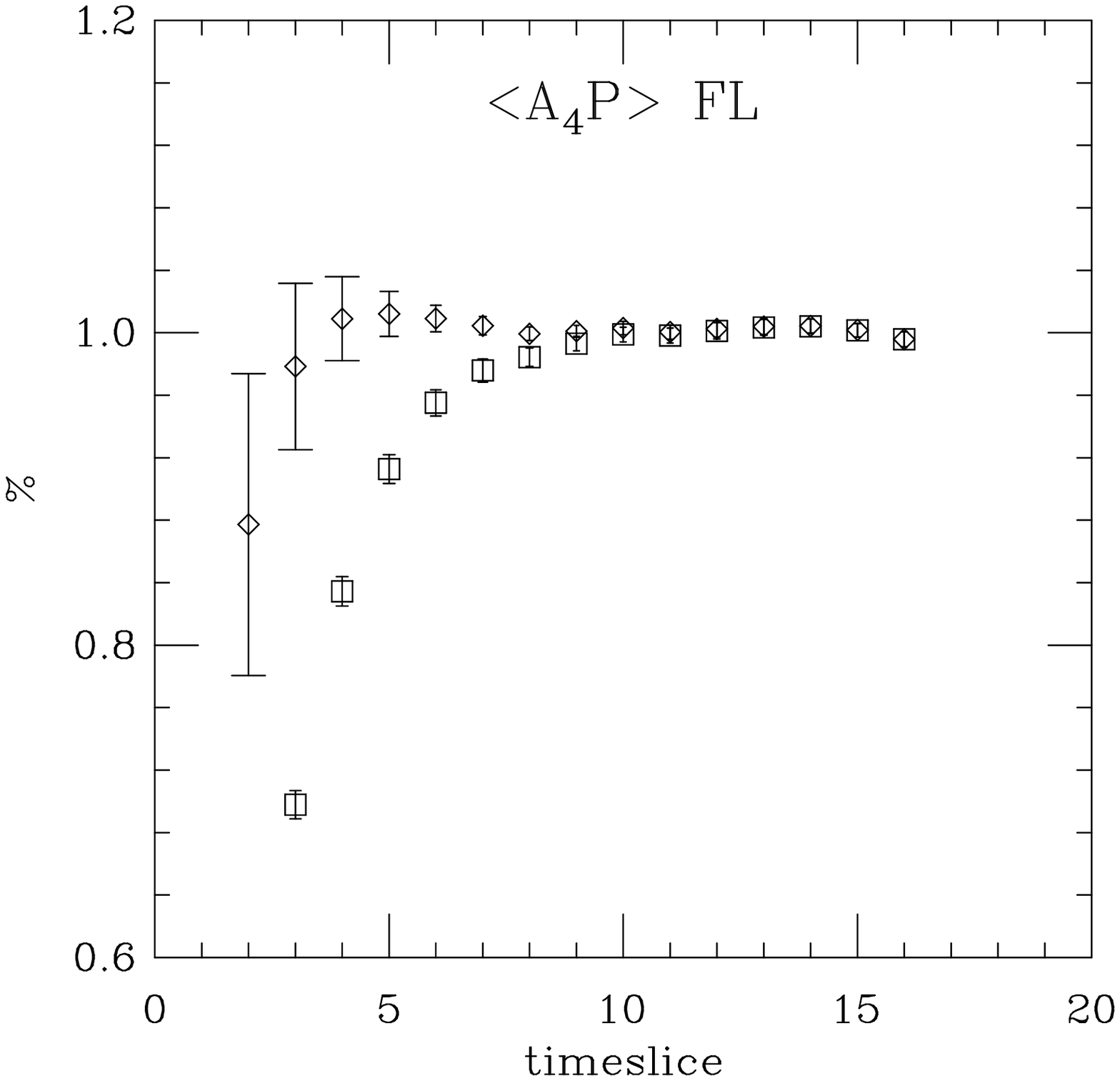}
\epsfxsize=8.0truecm\epsffile{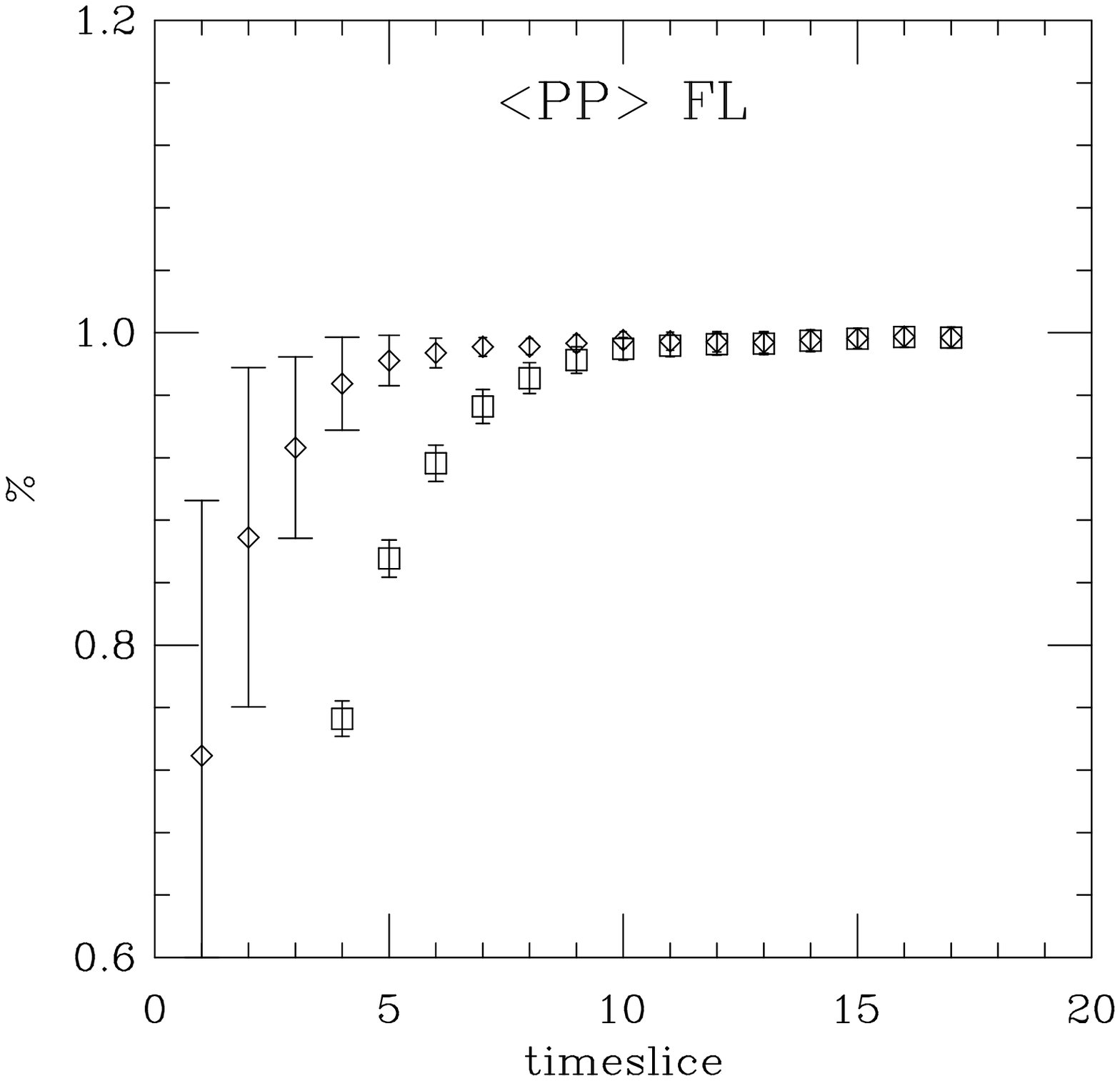}}
\caption{The fractional contribution of the ground state~(squares) and
the sum of the ground state and first excited state~(diamonds) to the
$C_{PA_4}$ and $C_{PP}$, $LL$ and $FL$ correlators for $\beta=5.93$
and $\kappa=0.1327$. }\label{fexp1}
\end{center}
\end{figure}

To implement the LANL method we first fix $t_{gs}$.
Figure~\ref{fexp1} shows the fractional contribution of the ground
state to the correlators which appear in $r$ and $s$ with
$\kappa_l=0.1327$ for the two types of correlators, $LL$ and $FL$,
that are available for this data set~(see
Appendix~\ref{multifits}). The ground state dominates by approximately
timeslice 12 for all correlators. We use $t_{gs}=12$ and as a check
also $t_{gs}=14$. $t_{ex}$ is allowed to vary in the region
$t_{ex}<t_{gs}$ while being careful to avoid any overlap with the
origin.

We present the details of the calculation of $c_A$ for the heaviest
$\kappa=0.1327$ using standard, $O(a^2)$, derivatives in
table~\ref{tone}.  The differences, $s(t_{ex})-s(t_{gs})$ and
$r(t_{ex})-r(t_{gs})$, are well determined for $t_{ex}$ close to the
origin, but rapidly decrease and become dominated by noise as $t_{ex}$
increases to around timeslice 8. The ratio of equation~\ref{calanl}
agrees with the results obtained using a fit apart from $t_{ex}=3$.
However, the LANL fit does not give a reasonable $Q$~(defined
as $Q>0.01$) until timeslice $5$ for $FL$
correlators and $7$ for $LL$ correlators.

The LANL method is expected to achieve good fits in a region where the
$O(a^2)$ errors in $m_{imp}(t)$~(equation~\ref{pcacimp}) are not
rapidly changing. In this region, the time dependence of $ac_As(t)$
can compensate for any corresponding variation in $r(t)$. This is
likely to occur when there are only a few states contributing to $r$
and $s$~\footnote{As $\beta$ increases and the $O(a^2)$ errors
decrease one expects a reasonable fit to be possible including more
excited states than on coarser lattices}. Ideally, in this region, the
estimates of $c_A$ are stable with a variation of $t_{ex}$. Any
dependence on $t_{ex}$, or any difference between the results from the
$LL$ or $FL$ correlators means that $O(a^2)$ contributions to $r$ and $s$
are appearing in $c_A$.

\begin{figure}
\begin{center}
\centerline{
\epsfxsize=8.0truecm\epsffile{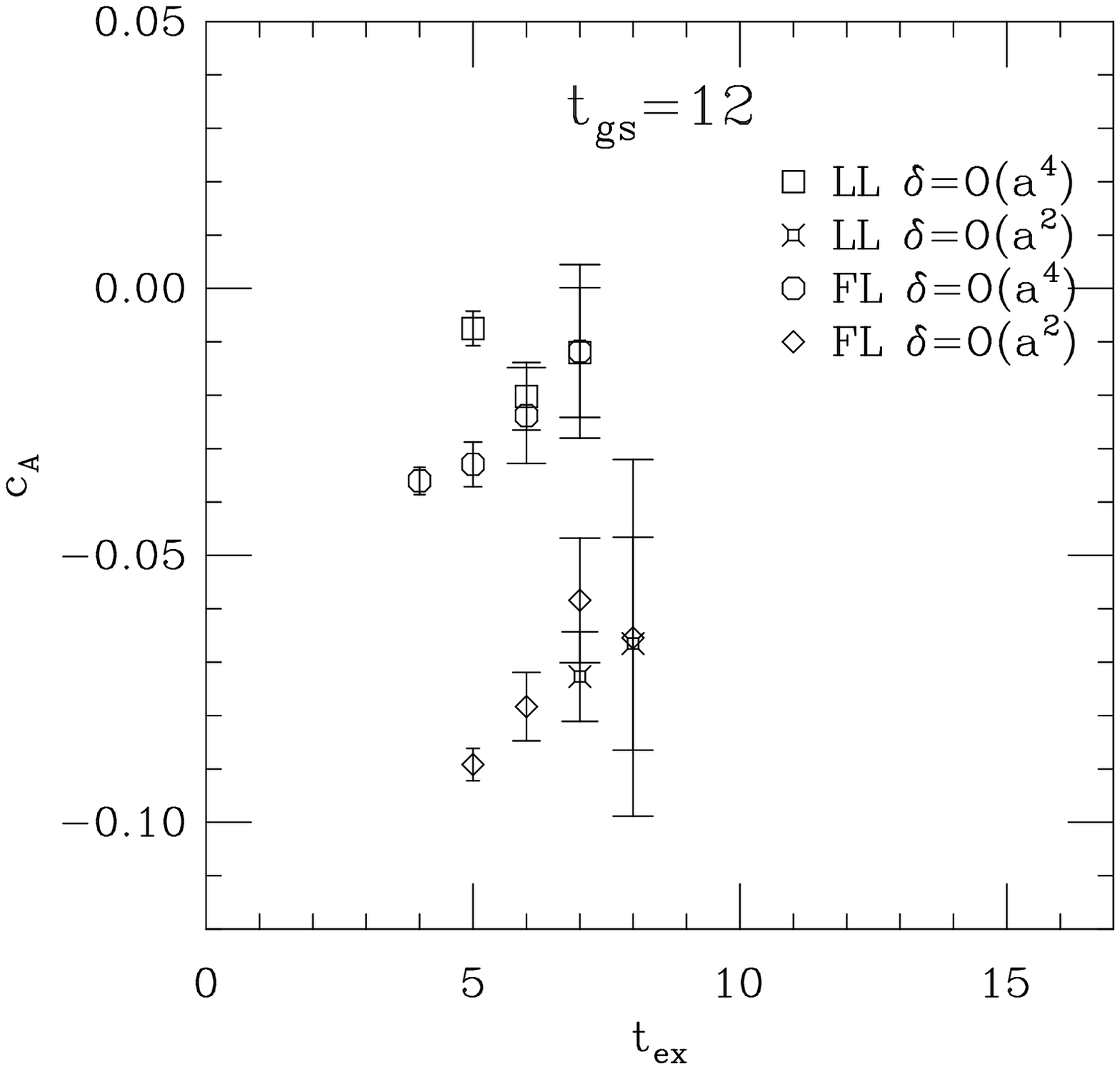}
\epsfxsize=8.0truecm\epsffile{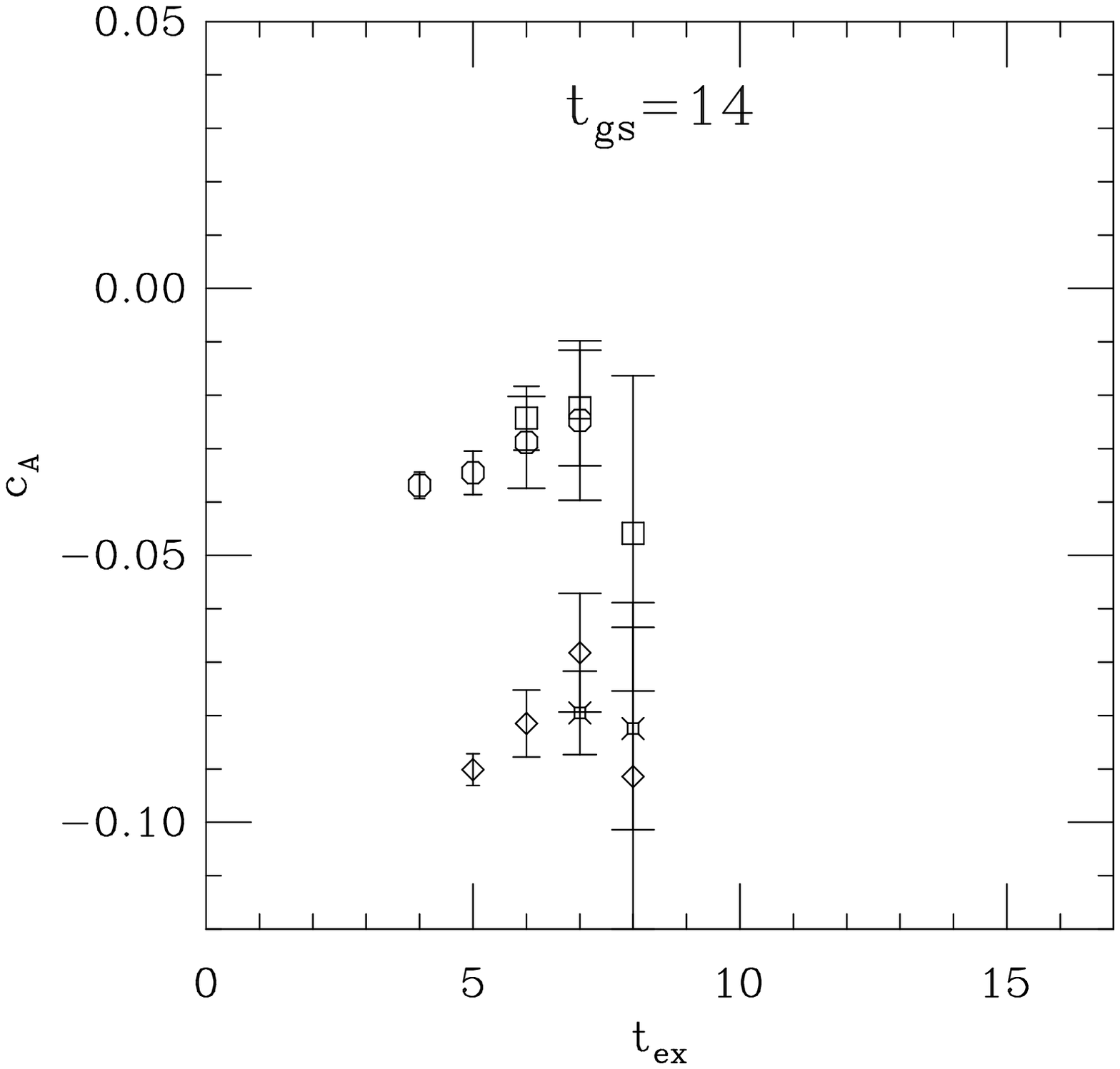}}
\caption{The values of $c_A$ obtained using the LANL method at $\beta=5.93$ and $\kappa_l=0.1327$ as a function of $t_{ex}$ used in the fit. $\delta$ indicates the type of lattice derivatives used.}\label{figt}
\end{center}
\end{figure}

The results from the LANL fits, with reasonable $Q$s, are shown in
figure~\ref{figt}. $c_A$ is stable with $t_{ex}$~(possibly excluding
$t_{ex}=5$ for $FL$ correlators), although noise rapidly dominates as
the fitting range becomes smaller. There is agreement between the
values obtained using $FL$ and $LL$ correlators, and $t_{gs}=12$ and
$14$. We can compare the range over which a good fit is found with the
fractional contribution of the sum of the ground and first excited
state to the correlators which make up $r$ and $s$, shown in
figure~\ref{fexp1}. Roughly, the earliest $t_{ex}$ with $Q>0.01$
corresponds to the timeslice when all but the first excited state and
the ground state dominate the correlators; since the $FL$ correlators
have a lower contribution from excited states, the ground state plus
first excited state dominate at an earlier timeslice compared to the
$LL$ correlators and a smaller $t_{ex}$ can be used.

\begin{figure}
\begin{center}
\centerline{
\epsfxsize=8.0truecm\epsffile{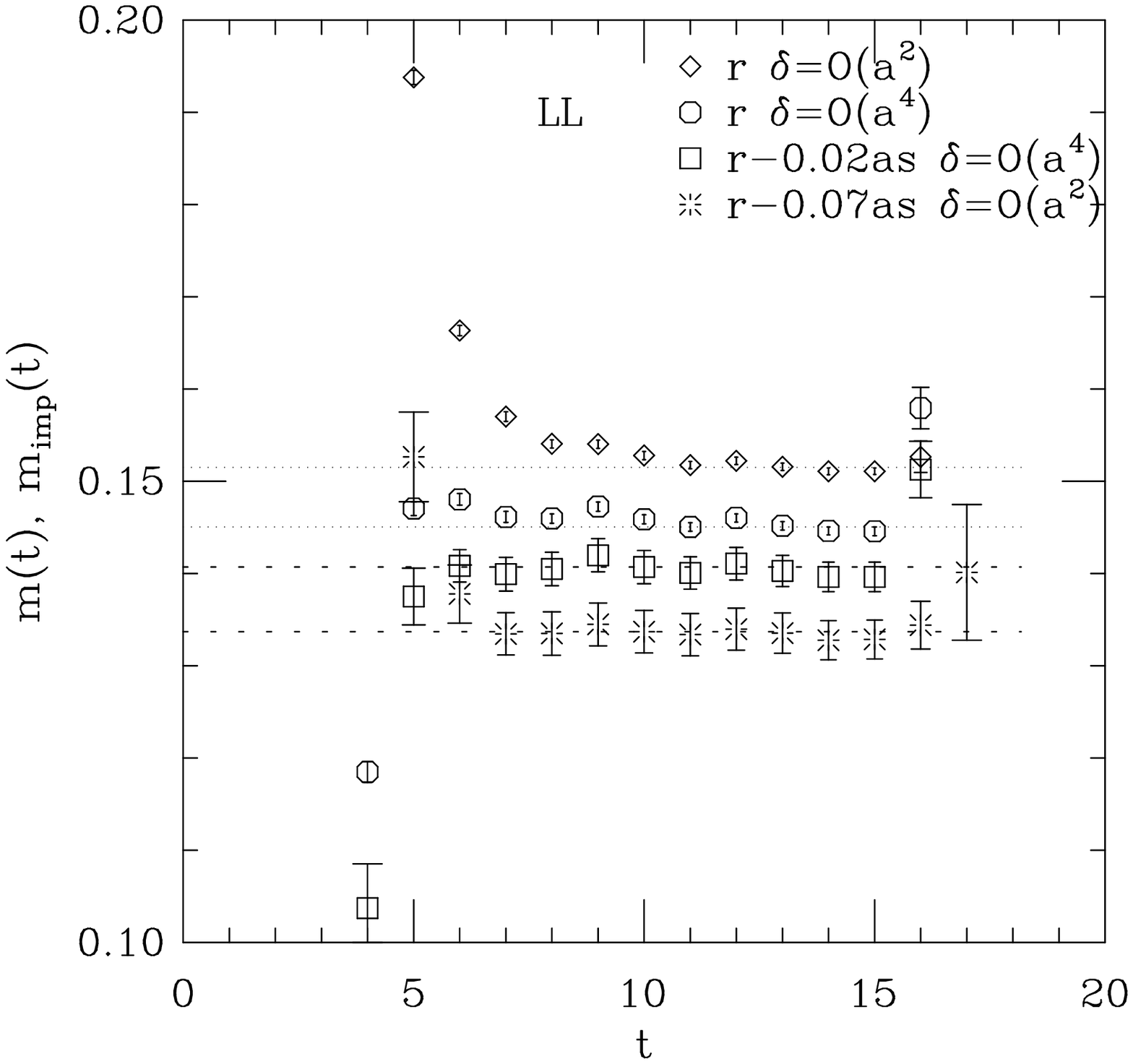}
\epsfxsize=8.0truecm\epsffile{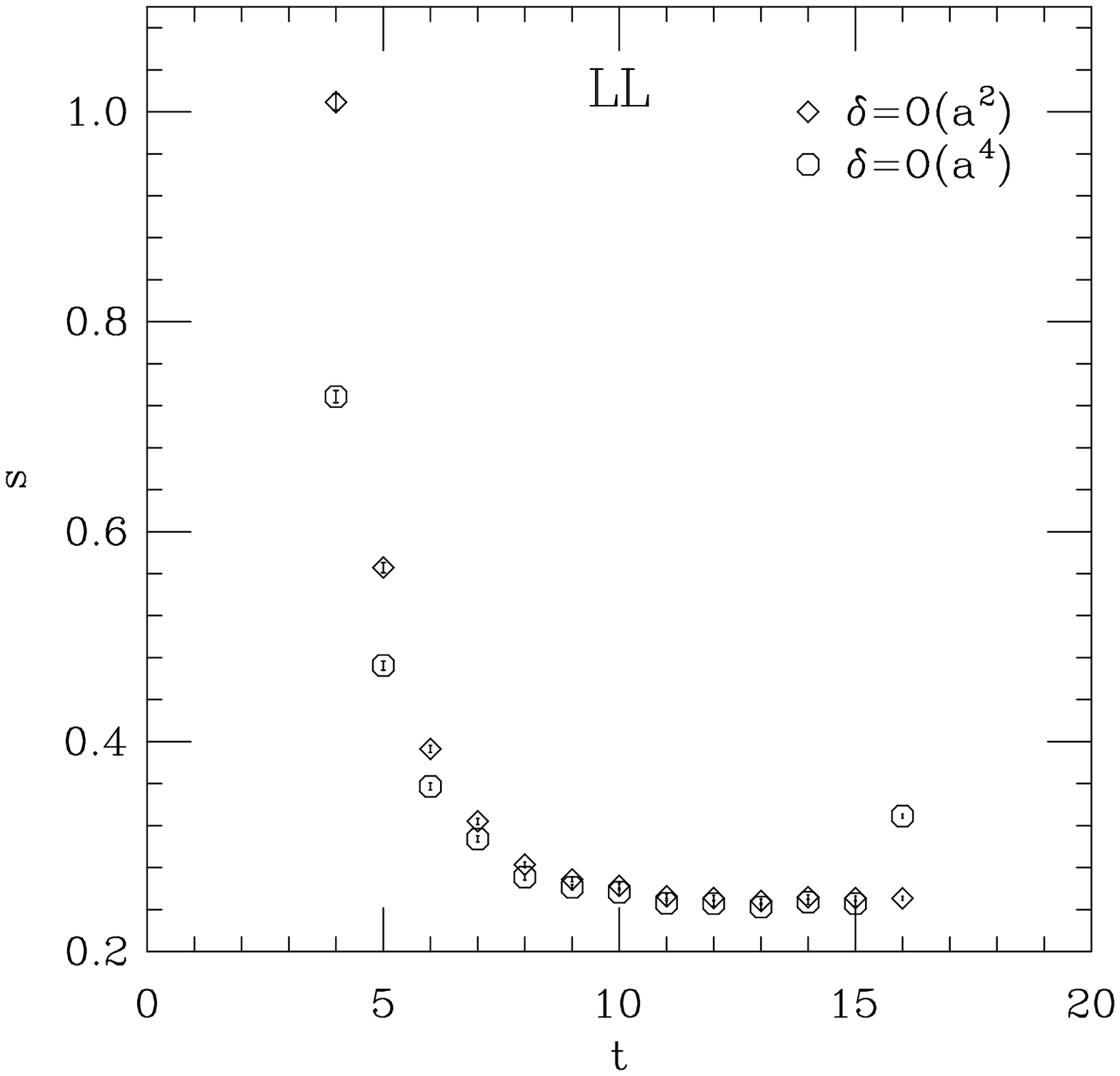}}
\caption{$m(t)=r(t)$, $s(t)$ and $m_{imp}(t)=r(t)+ac_As(t)$ as a 
function of timeslice at $\beta=5.93$ and $\kappa_l=0.1327$ for $LL$
correlators. The dotted lines indicate the values for $2am_{PCAC}$
obtained from fitting the results to a constant The dashed lines
indicate the value of $2am_{PCAC}^{imp}$ obtained from the fit for
$c_A$, where the fitting range $7-12$ was used to extract $c_A$ for
$O(a^2)$ derivatives~(bursts) and $6-12$ for $O(a^4)$~(squares). These
data points also include the statistical error of $c_A$.}\label{fig1}
\end{center}
\end{figure}

The results change quite significantly when we switch to improved
derivatives. Figure~\ref{fig1} shows the effects of changing the
derivatives on $m(t)=r(t)$ and $s(t)$ for $LL$ correlators. Clearly,
the time dependence of $m(t)$ is much reduced in the range $t=5-10$
when improved lattice derivatives are used, indicating that most of
the discretisation effects seen when using the standard derivatives
are due to the $O(a^2)$ error in $r(t)$ associated with this
derivative and not the $O(a)$ errors which we are trying to cancel
with $ac_As(t)$. A similar but much less dramatic effect is seen for
$s(t)$. This translates into much smaller values for
$r(t_{ex})-r(t_{gs})$, $s(t_{ex})-s(t_{gs})$ and $c_A$, as seen in
table~\ref{ttwo} and figure~\ref{figt}. 

For both $LL$ and $FL$ correlators, reasonable fits can be obtained
with slightly smaller $t_{ex}$ than with standard derivatives,
suggesting $\Delta r_J[O(a^2)]$ in this region is not large. As
figure~\ref{figt} shows, there is agreement to within $2\sigma$
between the $LL$ and $FL$ results, with the exception of the fit to
$LL$ correlators with $5-12$, which disagrees significantly with the
$FL$ result over the same fitting range. This is presumably due to a
larger $O(a)$ in $c_A$ for the $LL$ result, since these correlators
have a larger contribution from first excited~(and higher)
states. This fit is on the borderline of being considered reasonable;
changing $t_{gs}$ to $14$, the $Q$ drops further.

We demonstrate the effect of $O(a)$ improvement on $2am_{PCAC}$, for
various values of $c_A$ for $LL$ correlators in figure~\ref{fig1}.
The discretisation errors in $m(t)$ can be reduced using either
standard or improved lattice derivatives, however, the latter requires
a smaller value of $c_A$. As expected $m_{imp}(t)$ is constant over
the time range used in our fit, with the plateau being one
timeslice longer in the case of improved derivatives.

\begin{figure}
\begin{center}
\centerline{
\epsfxsize=8.0truecm\epsffile{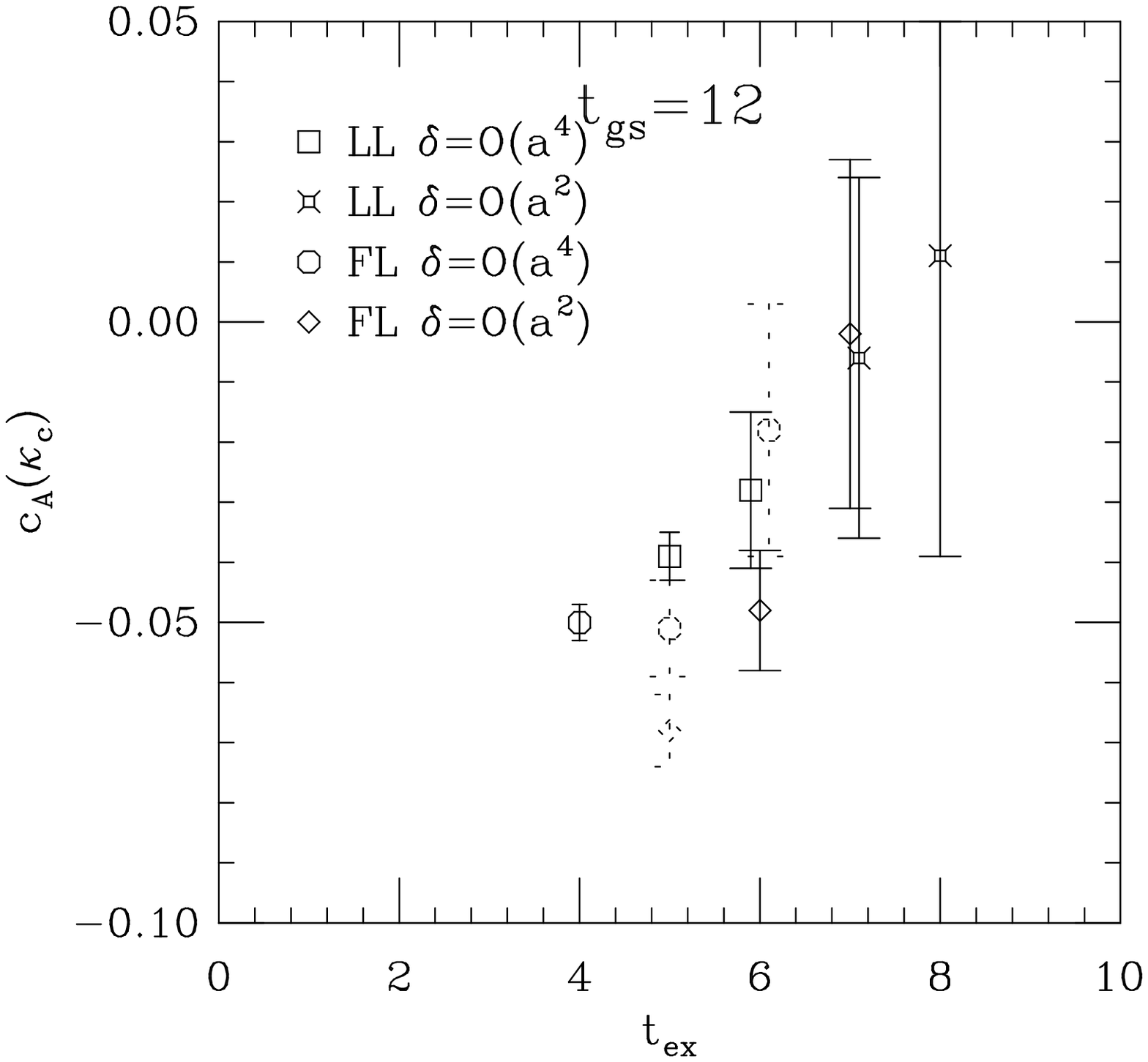}
\epsfxsize=8.0truecm\epsffile{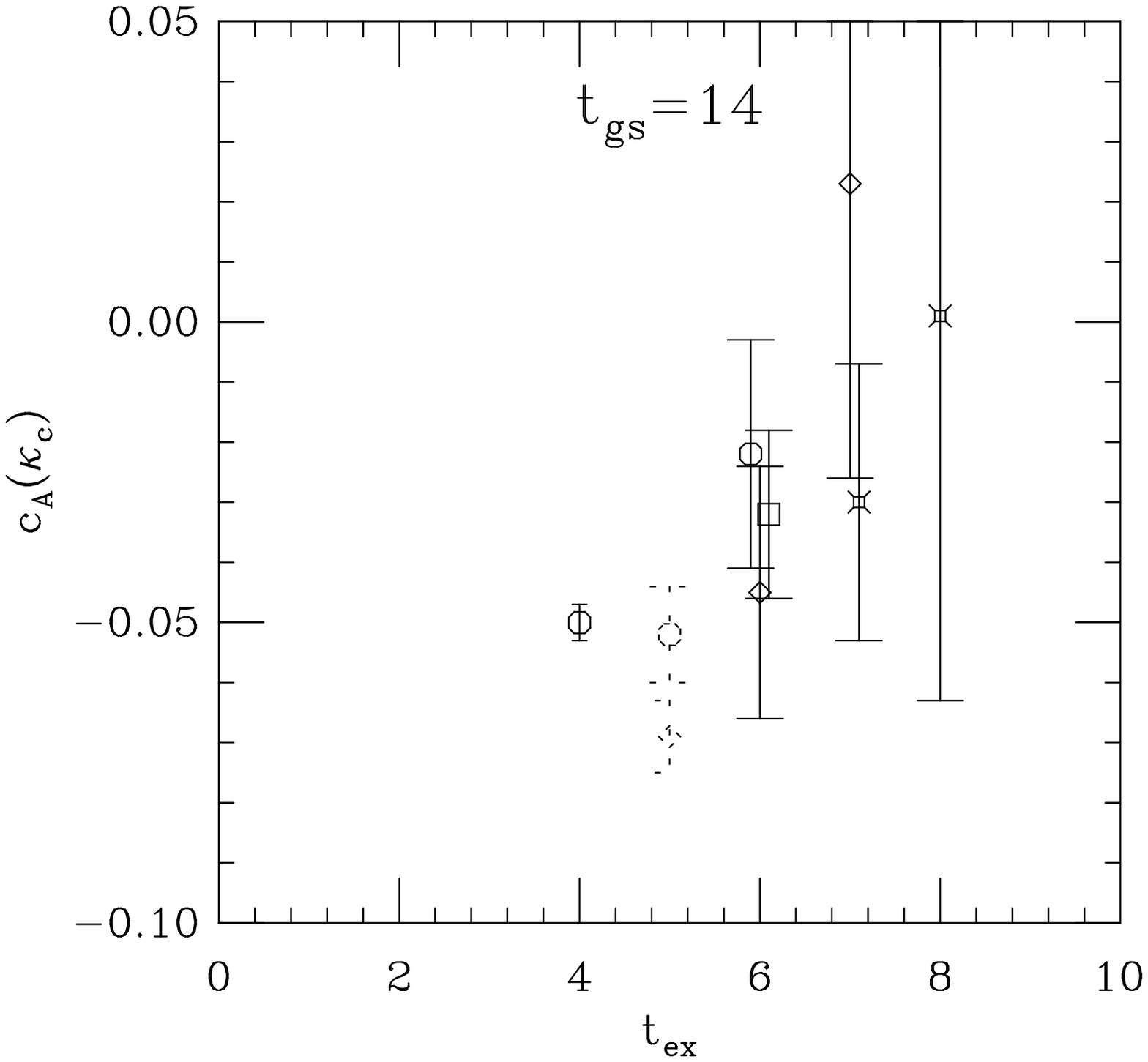}}
\caption{The values of $c_A$ obtained from chiral extrapolation 
in $aM_{PS}^2$ at $\beta=5.93$ as a function of $t_{ex}$ used in the
LANL fit. The dashed points indicate that an uncorrelated chiral
extrapolation was performed.}\label{figc}
\end{center}
\end{figure}

\begin{figure}
\begin{center}
\centerline{
\epsfxsize=8.0truecm\epsffile{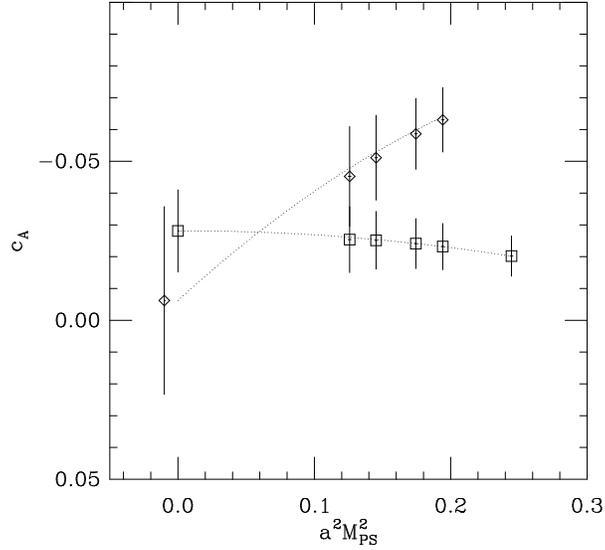}}
\caption{$c_A$ as a function of $aM_{PS}^2$ at $\beta=5.93$ for $LL$
correlators. The fitting range $6-12$ was used for $O(a^4)$
derivatives~(squares) and $7-12$ for $O(a^2)$
derivatives~(diamonds).}\label{figd}
\end{center}
\end{figure}

Our improvement condition is defined in the chiral limit and the
results for $c_A$ must be extrapolated to zero quark mass. Details of
this procedure are given in Appendix~\ref{secchiralca} and the
results are presented in figure~\ref{figc} and in
table~\ref{chiralttwo}. In general, $c_A$ from using standard
derivatives has a bigger statistical error than that obtained using
improved derivatives since $t_{ex}$ is larger and the chiral
extrapolation is usually more severe. The latter point is illustrated
in figure~\ref{figd}. From the discussion above, the large value of
$c_A$ from standard derivatives is due to a large contribution to
$m_{imp}(t)$ from the $O(a^2)$ errors associated with these
derivatives.  These errors are $m_q$ dependent and should disappear in
the chiral limit. The figure shows that $c_A$ drops significantly with
quark mass, agreeing with the result from the improved derivatives in
the chiral limit. In contrast $c_A$ from improved derivatives is much
less dependent on the quark mass.

Comparing all results in figure~\ref{figc} we find $c_A(\kappa_c)$
from different derivatives, smearing and from different $t_{ex}$ and
$t_{gs}$ are consistent in the chiral limit~(with the exception of
$t_{ex}=5$ for $FL$ correlators). We also implemented the LANL lattice
derivatives~\cite{lanl} and found consistent results in the chiral
limit. $c_A=-0.032(14)$ from $LL$ correlators, fitting $6-14$, using
improved derivatives is taken as the final value for $c_A$ at this
$\beta$. The error reflects the spread in values for $c_A$ and
indicates the uncertainty from some of the associated $O(a)$ effects.

We now consider applying the $LANL$ method at different $\beta$s.  In
principle, one should ensure, as accurately as possible, that the same
improvement conditions are applied to determine $c_A$ as $\beta$ is
changed. This ensures the systematic errors are correlated between
different determinations and $c_A$ smoothly extrapolates to zero in
the continuum limit. For example, we need to keep the proportion of
the excited states to ground state contributing to $r(t_{ex})$
fixed. This requires a tuning of the smearing function~\footnote{It is
advantageous to work in the regime where the first excited state is
the only significant radial excitation. A smearing with a good overlap
with this state~(likewise for another smearing with the ground state)
would enable this proportion to be fixed accurately as $\beta$
changes.}, which was not possible in this study~(and was not attempted
in reference~\cite{lanl2}). Thus, we have chosen a fairly conservative
error for $c_A$ to take into account the difficulty in applying the
same improvement conditions for the other simulations. In principle
any value of $c_A$ in figure~\ref{figc} is a valid estimate of the
coefficient for a particular simulation. A more aggressive choice
for $c_A$ would be, for example $-0.050(3)$ from $FL$ correlators with
$t_{ex}=4$. In the following, to keep the systematic errors as
correlated as possible we extract final numbers for $c_A$ from $LL$
correlators and improved lattice derivatives~(as used for our choice
of $c_A$ above).

In addition, the physical volume of the simulation should also be kept
fixed when determining $c_A$. We comment on this in the next section.

\subsection{Results at $\beta=6.0$ and $\beta=6.2$}

\begin{figure}
\begin{center}
\centerline{
\epsfxsize=8.0truecm\epsffile{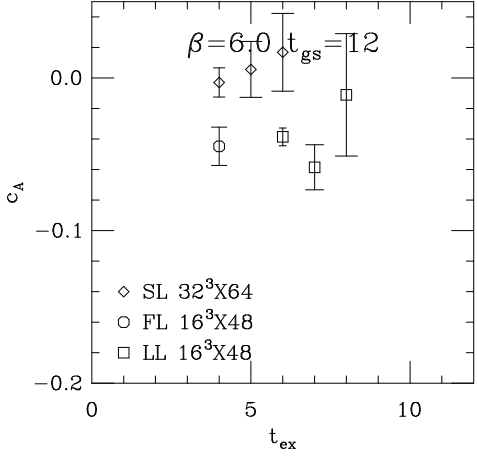}
\epsfxsize=8.0truecm\epsffile{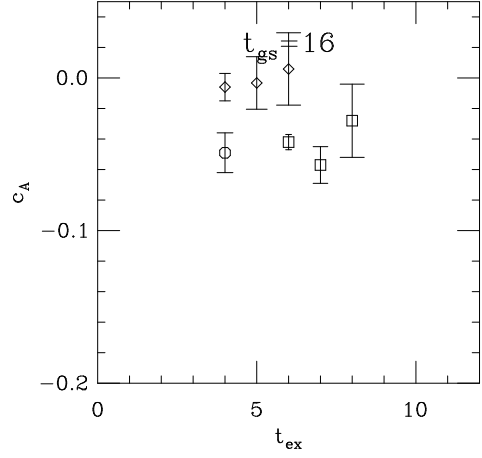}
}
\centerline{(a)\hspace{10cm}(b)}
\centerline{
\epsfxsize=8.0truecm\epsffile{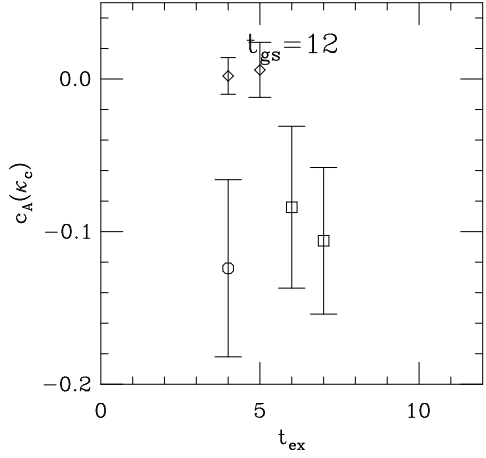}
\epsfxsize=8.0truecm\epsffile{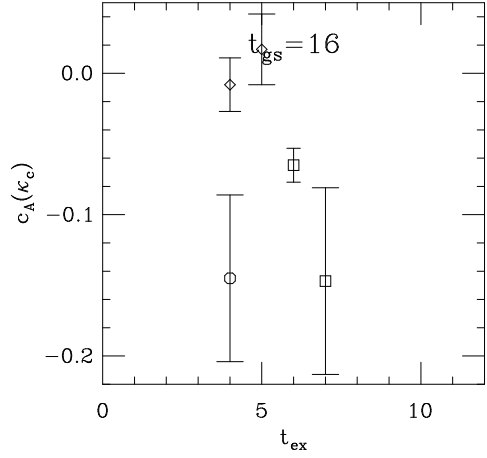}
}
\centerline{(c)\hspace{10cm}(d)}
\caption{(a) $c_A$ as a function of $t_{ex}$ for
$\beta=6.0$ and $\kappa_l=0.13344$ with $t_{gs}=12$. $O(a^4)$ improved
derivatives have been used.  We found that the $Q$s for the LANL fits
were higher than those obtained at $\beta=5.93$. (b) the same as in
(a) but with $t_{gs}=16$.  (c) $c_A$ in the chiral limit from
$t_{gs}=12$. (d) $c_A(\kappa_c)$ from $t_{gs}=16$.}\label{fig2}
\end{center}
\end{figure}

Considering the analysis at $\beta=6.0$ first, we present the results
for $c_A$ in figure~\ref{fig2} and table~\ref{chiraltthree}. In
addition to the simulations on the $16^3$ volume with $FL$ and $LL$
correlators, $c_A$ was also calculated on a small ensemble of large
volume~($32^3$) configurations with $SL$ smearing~(unfortunately no
$LL$ correlators were available). As discussed in
Appendix~\ref{multifits}, the fuzzed smearing was optimised for the
ground state and hence the first excited state amplitude for the $FL$
correlators is very small; an estimate for $c_A$ could only be
extracted for one value of $t_{ex}$. Nevertheless, the $FL$ and $LL$
results are consistent and there is no significant variation with
$t_{gs}$. 

A discrepancy was found, however, when comparing to the $SL$ results
on the larger volume. This can be seen in figure~\ref{fig2} for $c_A$
at finite $m_q$ around the strange quark mass.  There is a $3.5\sigma$
disagreement between the $FL$ value for $t_{ex}=6$ and the $SL$ value
at $t_{ex}=4$~($t_{g.s.}=16$).  The discrepancy between the results
could be due to $O(a)$ terms arising from the use of different
smearings or it may indicate a finite volume effect. The $16^3$
lattice corresponds to a physical volume of approximately
$(1.5\mbox{fm})^3$, which is probably too small to accommodate excited
pseudoscalar states. $c_A$ itself is an ultra-violet quantity but may
be affected by finite volume effects because of the matrix elements
being used to determine it.

We attempted to investigate finite volume effects by comparing masses
and decay constants~(the matrix element $<P\partial_4A_4>$, in $r(t)$,
is related to $f_{PS}$) on the two volumes. Unfortunately, we were
only able to extract these quantities for the ground state on the
large volume. We found an $.8\%$ or $2.5\sigma$ decrease in $m_{PS}$
changing from the small to large volume and no significant change in
$f_{PS}$~(this is in agreement with the results in
reference~\cite{ukqcd}). A more thorough investigation of finite
volume effects is needed. If the physical volume was the same as at
$\beta=5.93$, it would not matter how dependent $c_A$ is on the size
of the lattice, since it is a higher order effect. However, the finer
lattice is $16\%$ smaller than that at $\beta=5.93$. This must be
considered when quoting an error on $c_A$.

The chiral extrapolations of at $\beta=6.0$ proved difficult for $c_A$
from the $LL$ and $FL$ correlators. Apart from the $LL$ result for
$t_{ex}-t_{gs}=6-16$, the errors on the extrapolated values are very
large due to having to use a fit function quadratic in $a^2M_{PS}^2$
and/or only being able to fit to the lightest data points. The
discrepancy of the $6-16$ result with the $4-16$ $SL$ value is
approximately $2.5\sigma$.

As noted in the previous section, to keep the same improvement
conditions $c_A$ should be extracted using a $t_{ex}$ with the same
relative proportion of ground state to excited states as that used at
$\beta=5.93$. One possibility, concentrating on $LL$ correlators, is to
fix $t_{ex}$ to correspond to the same physical time; $t_{ex}=6$
chosen at $\beta=5.93$ corresponds to approximately timeslice $7$ for
$\beta=6.0$. The statistical errors of $c_A(\kappa_c)$ for this
$t_{ex}$ are too large for the estimate to be useful in our later
analysis. If we choose $t_{ex}=6$ then the errors do not reflect the
unresolved $O(a)$ or finite volume effects mentioned above. These
problems motivate us to discard the results for $c_A$ at this $\beta$.

\begin{figure}
\begin{center}
\centerline{
\epsfxsize=8.0truecm\epsffile{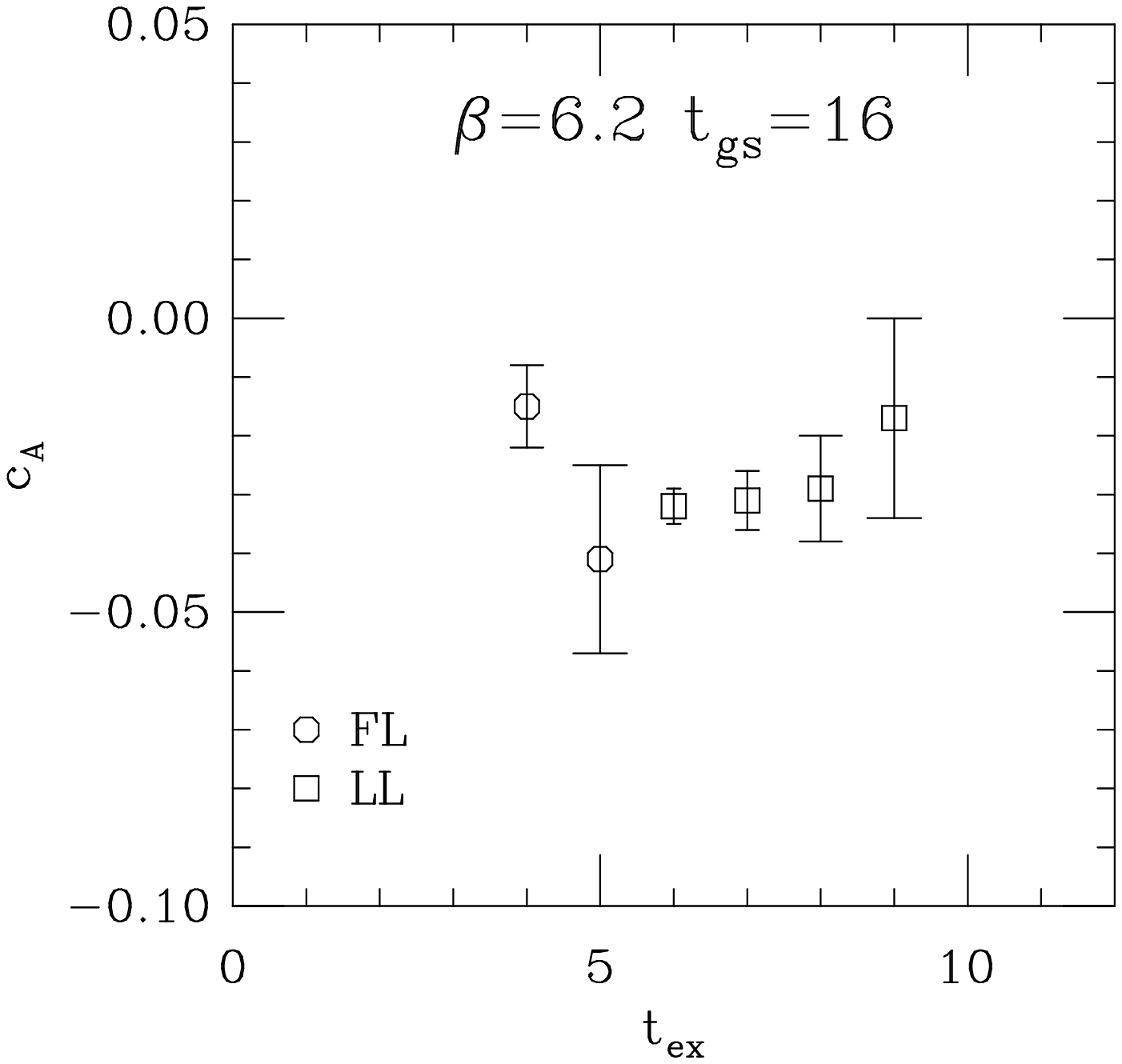}
\epsfxsize=8.0truecm\epsffile{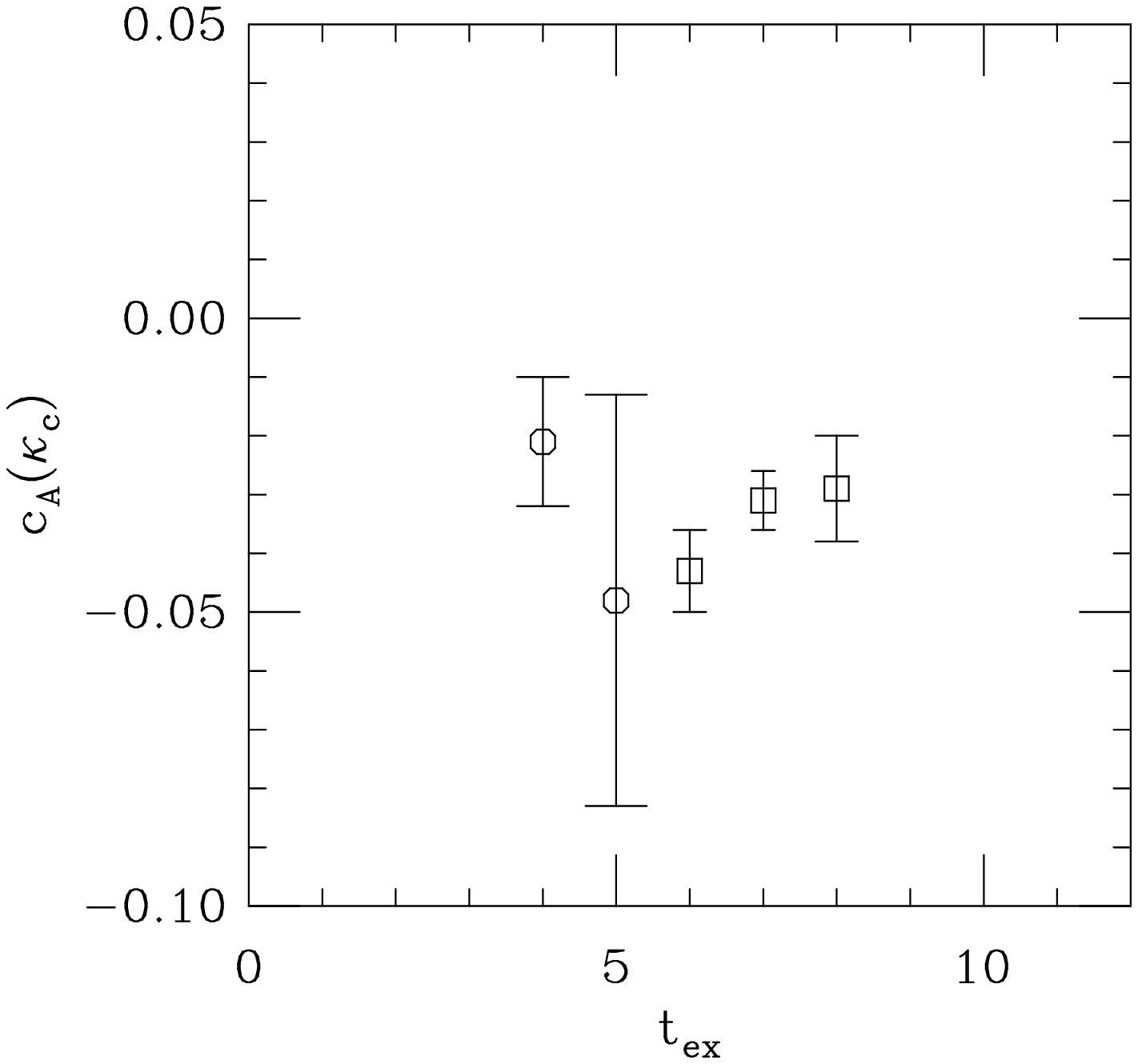}
}
\centerline{(a)\hspace{10cm}(b)}
\caption{(a) $c_A$ as a function of $t_{ex}$ for
$\beta=6.2$ and $\kappa=0.13460$ with $t_{gs}=16$. $O(a^4)$
improved derivatives have been used. (b) $c_A$ in the chiral
limit.}\label{fig2b}
\end{center}
\end{figure}

At $\beta=6.2$ the situation is more straightforward as displayed in
figure~\ref{fig2b} and table~\ref{chiraltfour}. There is consistency
between the results from different smearings, where the fuzzed
smearing is optimised in a similar way to that at $\beta=6.0$, and
also as $t_{ex}$ is varied. We also found no change in the results if
$t_{gs}$ is increased and there was no difficulty with chiral
extrapolation. Keeping the same physical $t_{ex}$ as at $\beta=5.93$
corresponds to using timeslice $9$. Unfortunately, the estimates of
$c_A$ have fallen into noise at this point. However, given the
stability of the results with $t_{ex}$ we are unlikely to introduce
significant systematic errors if we choose a fitting range of
$7-16$. Thus, our final result at this $\beta$ is $c_A=-0.031(5)$.
The lattice volumes at $\beta=5.93$ and $6.2$ are fairly close in
physical size, and we assume that the error on $c_A$ is sufficient to 
compensate for the small discrepancy.

\subsection{Comparison with previous results}
Figure~\ref{falpha} compares our results with those of the ALPHA
collaboration~\cite{alpha} and the LANL group~\cite{lanl,lanl2} in the
range of $\beta$s we have simulated.  Unfortunately, our errors on
$c_A$ are quite large after chiral extrapolation. Nevertheless, we
obtained consistency with the LANL results, in particular at coarser
lattice spacings. The LANL results are split up into those extracted
using standard $O(a^2)$ derivatives and those obtained using modified
$(a^2)$ derivatives, mentioned in section~\ref{capcac}. The latter is
closer to the choice of derivatives employed here and we find greater
consistency with our results, at $\beta=6.2$, in this case. The LANL
results have smaller errors compared to our values even though our
study has much higher statistics. We believe this is due in part to
our more conservative error estimates to take into account the
difficulty in applying the improvement conditions consistently as
$\beta$ changes. In addition, the LANL group employed much heavier
light quark masses, over a wider range, than in our analysis, and this
led to lower statistical errors after chiral extrapolation.

The results from the LANL method are very slowly varying with $\beta$
and do not change significantly from $\beta\sim 6$ to $6.2$. This is
in agreement with the 1-loop perturbative results, also shown in the
figure for $\alpha=\alpha_P(1/a)$~\cite{alphap}, where $c_A =
-0.0952\alpha$~\cite{ca,lanl2}.  Our results, with large errors, are
consistent with the perturbative result, however the LANL values are
somewhat higher. The uncertainty in the perturbative value, from
higher order terms, is difficult to estimate. One could take anything
between the square of the 1-loop term, $\delta c_A\sim 0.0006$ to
$1\alpha^2\sim 0.04-0.08$ in the range of $\beta=6.2$ to $5.93$. The
2-loop contribution would have to be quite large to obtain consistency
with the LANL results~\cite{lanl2}. However, the calculation of this
contribution would significantly reduce the uncertainty on $c_A$; the
perturbative result is valid in the infinite volume limit and there is
no $O(a)$ ambiguity in $c_A$~\footnote{Although, of course,
$O(\alpha^n a)$ terms remain in the axial-vector current.} which is
present in the nonperturbative determination and can be large, in
particular at coarser lattice spacings.

This can be seen when comparing our results~(and those of the LANL
group) with those of the ALPHA collaboration. At $\beta=6.2$, all
results are consistent, however, at coarser lattice spacings a large
discrepancy appears as the ALPHA $c_A$ rapidly increases~\footnote{At
$\beta=5.93$, $c_A=-0.16$ from the Pade expansion of the ALPHA
results~\protect\cite{alpha}. However, it is more likely to be around
$-0.11$~\protect\cite{hartmut}.}. This discrepancy indicates how large
the $O(a)$ ambiguity in $c_A$ can be. In addition, the LANL group
using new results at $\beta=6.4$ have found that the difference
between their results and that of the ALPHA collaboration requires
$O(a^2)$ terms as well as $O(a)$~\cite{rajan}. We believe that by
extracting $c_A$ in a region where only the first excited state and
ground state contributes, looking for consistency between results from
different smearings and using improved derivatives we have minimized
the $O(a)$ artifacts in $c_A$ within the LANL method.
Nonetheless, if the improvement conditions are kept fixed accurately
as $\beta$ is changed, large variations in the estimates of $c_A$ are
not important. However, practical difficulties arise if the choice
taken leads to significantly worse scaling violations for physical
predictions, namely, $f_\pi$ and the renormalised quark mass.

\section{Scaling of $f_\pi$ and $m^{\overline{ms}}$}
\label{scalingsec}
$C_A$ is needed, for $O(a)$ improved estimates of the pseudoscalar
decay constant and the quark mass determined from the bare PCAC quark
mass.  Figure~\ref{fig2e} displays our results for the renormalised
decay constant in units of $r_0$ as a function of the squared meson mass
using our values for $c_A$, at $\beta=5.93$ and
$6.2$ and also using $c_A$ determined by the ALPHA collaboration
at $\beta=6.0$ and $6.2$. The extraction of the decay constant and the
renormalisation factors used are detailed in
Appendices~\ref{multifits} and~\ref{renormfandm}.

The figure clearly shows that using our smaller values of the
improvement coefficient there are no significant scaling violations
between $\beta=5.93$ and $\beta=6.2$, in contrast to the significant
violations found using the ALPHA values for $c_A$. In the same figure
we plot $r_0f^{ren}$ as a function of $(a/r_0)^2$ for the reference
mass $(r_0M_{PS})^2=3.0$. To see whether both data sets are consistent
with the same continuum limit we fit the data simultaneously with a
function linear in $(a/r_0)^2$ and obtain a $\chi^2/d.o.f.=2.4$.  This
is on the borderline of being considered a good fit. Results at more
$\beta$ values are needed in order to be able to include higher order
terms in the continuum extrapolation and accommodate the results
obtained using the ALPHA values of $c_A$.

We calculated the renormalised quark mass from the improved bare PCAC
quark mass. Details of this calculation and a consistency check,
extrapolating $(r_0M_{PS})^2$ as a function of $r_0m_{PCAC}^{imp}$,
are given in Appendices~\ref{multifits} and~\ref{renormfandm}. We note
that in order for the meson mass and the PCAC mass to vanish at the
same point, chiral log terms had to be included in the chiral
extrapolation. Similarly, we found that when these terms are included
to extract $\kappa_c$ from $M_{PS}^2$, consistency is obtained with
$\kappa_c$ extracted from $m_{PCAC}^{imp}$.

Figure~\ref{fig2f} presents our results for the renormalised quark
mass in the $\overline{MS}$ scheme at the scale $2$~GeV as a function
of $(a/r_0)^2$ for the reference mass $(r_0M_{PS})^2=3.0$. In (a) the
calculation proceeds via the renormalisation-group invariant mass
using nonperturbative renormalisation~\cite{squark} and the overall
statistical and systematic errors are small. Again, the scaling
violations are small using our values of $c_A$, in contrast to the
severe lattice spacing dependence seen using the ALPHA values. A
combined linear fit to both data sets has a rather poor
$\chi^2/d.o.f.=3.3$. As for the decay constant, more points are needed
to check for a consistent continuum limit.

In (b) we compare $m^{\overline{MS}}$ determined from
$\kappa_c$~(which is independent of $c_A$) with that obtained from the
PCAC mass. 1-loop perturbation theory is used for the renormalisation
factors and hence the overall errors are much larger than in (a). The
best scaling behaviour is seen for $m^{\overline{MS}}_{\kappa_c}$,
followed by $m^{\overline{MS}}_{PCAC}$ determined using our values of
$c_A$. A linear fit to the combined data gives, $\chi^2/d.o.f.=0.1$,
but the errors are rather large.

\section{Conclusions}
\label{conc}

We have undertaken a study of the difficulties and systematic errors
inherent in a nonperturbative determination of $c_A$ using the LANL
method. Some of these points may also apply to determinations using
the ALPHA method.

For practical purposes it is desirable that the determination of $c_A$
is defined so that the improvement term for the axial-vector current
removes only $O(a)$ errors and does not unnecessarily add large
additional $O(a^2)$ artefacts. We conclude that this is possible with
the LANL method but careful tuning of the improvement condition at
each $\beta$ is required so that physically the same condition is
imposed. Chiral extrapolations and finite volume effects can also
cause problems.

Our values of $c_A$, even on ensembles with high statistics, are
rather imprecise after these considerations but improved errors would
be possible with better smearings. Nevertheless it is clear that the
smaller values of $c_A$ that we obtain, compared to those of the ALPHA
collaboration, give improved scaling of the pion decay constant and the
renormalised quark mass.

\section{Acknowledgments}
We thank T.Bhattacharya and R.Gupta for discussions on the details of
their method. We also appreciated useful discussions with C.~Maynard,
R.~Sommer and H.~Wittig. S.~Collins has been supported by a Royal
Society of Edinburgh fellowship. This work was supported by the
Particle Physics and Astromony Research Council~(PPARC) through grants
GR/L56336 and GR/L2997, the European Community's Improving Human
Potential Programme under contract HPRN-CT-2000-00145, Hadrons/Lattice
QCD and the US Department of Energy~(DOE) under grant
DE-FG02-91ER40690.
\begin{appendix}
\section{Simulation Details}

The simulation details of the UKQCD quenched configurations are
compiled in table~\ref{sim}. Light quark propagators were generated
using the $O(a)$ improved Wilson action with the nonperturbative value
of the improvement coefficient $c_{SW}$ determined by the ALPHA
collaboration~\cite{alpha} at $\beta=6.0$ and $6.2$ and the SCRI
collaboration~\cite{scri} at $\beta=5.93$. In both cases the
interpolating curves which were found to fit the nonperturbative
determinations of $c_{SW}$ were used~(rather than, for example, the
numerical results at $\beta=6.0$ and $6.2$). The values of $\kappa$
were chosen in order to straddle the strange quark mass. The quark
propagators were tied together to form mesons in both degenerate and
non-degenerate $\kappa$ combinations.

Gauge-invariant smearing was applied at the source and/or sink using
either extended ``fuzzed'' spatial functions~\cite{fuzzing}~(denoted
$F$) or Gaussian-like spatial functions using the Jacobi
method~\cite{jacobi}~(denoted $S$). In the table, $FL$ refers to a
meson made up of a light quark propagator with a fuzzed source and
local sink, combined with a $LL$ antiquark propagator.
%
%

A previous analysis of the data sets at $\beta=6.0$ and
$6.2$~\cite{ukqcd} identified a small number of exceptional
configurations: 3 on the small volume at $\beta=6.0$, 2 for the large
volume and 1 configuration at $\beta=5.93$. These configurations were
removed from the statistical ensemble. For more details
see~\cite{ukqcd}.  The lattice spacing is set using $r_0=0.5$~fm and
the interpolating curve determined by the ALPHA collaboration for
$r_0/a$~\cite{r0}, from which we obtain $r_0/a=4.741$, $5.368$ and
$7.360$ at $\beta=5.93$, $6.0$ and $6.2$ respectively. Throughout the
analysis we ignore all errors in $r_0$.

\section{Extracting masses and decay constants}
\label{multifits}
We extract the mass and decay constant of the pseudoscalar meson from
the two-point correlation functions,
\begin{eqnarray}
C_{P A_4} & = & \left<P^\dagger(0)A_4(t)\right>\\
C_{P P} & = & \left<P^\dagger(0)P(t)\right>
\end{eqnarray}
where, 
\begin{eqnarray}
C_{P A_4} & = & \sum_{n=0}^{\infty} \frac{1}{2M_{P}^n} \left(\left<0|P^\dagger|n\right>\left<n|A_4|0\right>e^{-M_{P}^n t} - \left<0|P^\dagger|n\right>\left<n|A_4|0\right>e^{-M_{P}^n(T-t)}\right)\\
    & = & \sum_{n=0}^{\infty} A_{PA_4}^ne^{-M_P^nT/2}\sinh\left(M_P^n(t-\frac{T}{2})\right)\label{cap}\\
C_{P P} & = & \sum_{n=0}^{\infty} A_{PP}^ne^{-M_P^nT/2}\cosh\left(M_P^n(t-\frac{T}{2})\right) \label{cpp}
\end{eqnarray}
and $M_P^n$ is the mass of the nth radial excitation of the
pseudoscalar meson.  At $\beta=6.0$ and $6.2$, we performed a
simultaneous correlated fit to 5 correlators, $C_{PA_4}^{FL}$,
$C_{PP}^{FL}$, $C_{PP}^{FF}$, $C_{PA_4}^{LL}$ and $C_{PP}^{LL}$, where
$FL$ etc refers to the smearing of the correlator.  This combination
was chosen in order to extract information on the excited states as
well as the ground state masses and amplitudes. To increase statistics
the correlators were averaged about the center of the lattice.

Fits including the ground, first and second excited states were
attempted. Obtaining these fits at $\beta=5.93$ was straight
forward. However, at $\beta=6.0$ the fits lie consistently below
the central values of the 5 correlators, over any given fitting range,
even with $Q>0.1$. This is due to the fact that the small eigenvalues
of the covariance matrix for the fits are not determined well enough
with the statistics we have available. However, it is these
eigenvalues which dominate the minimization of the $\chi^2$. We chose
to zero eigenvalues which where below a certain cut-off,
$c_{cut}w_{max}$, where $w_{max}$ is the largest eigenvalue and
$c_{cut}=10^{-3}$ for all $\kappa$ combinations. This resulted in
removing approximately $16$~($22$) eigenvalues for a 1~(2) state fit.

\begin{figure}
\begin{center}
\centerline{
\epsfxsize=8.0truecm\epsffile{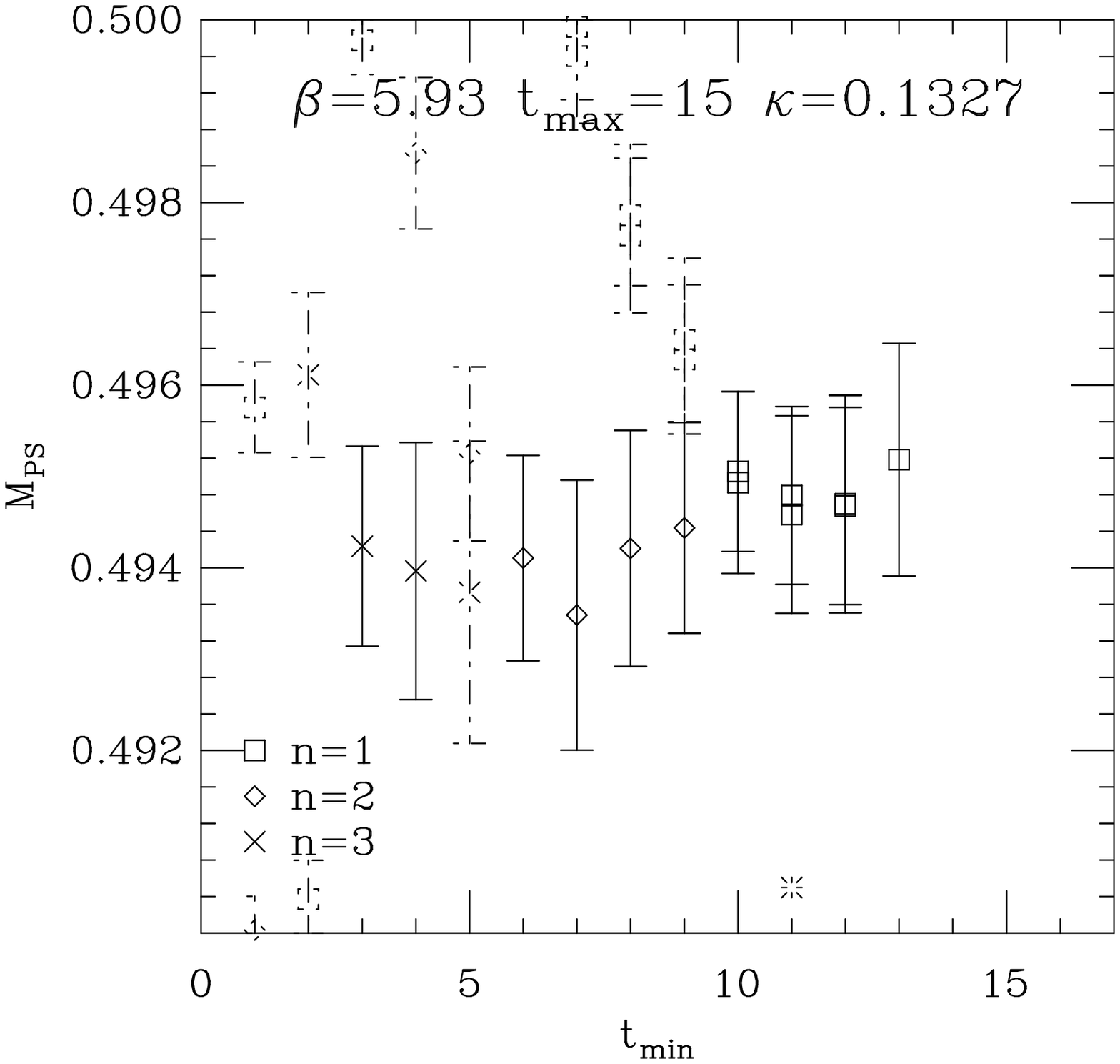}
\epsfxsize=8.0truecm\epsffile{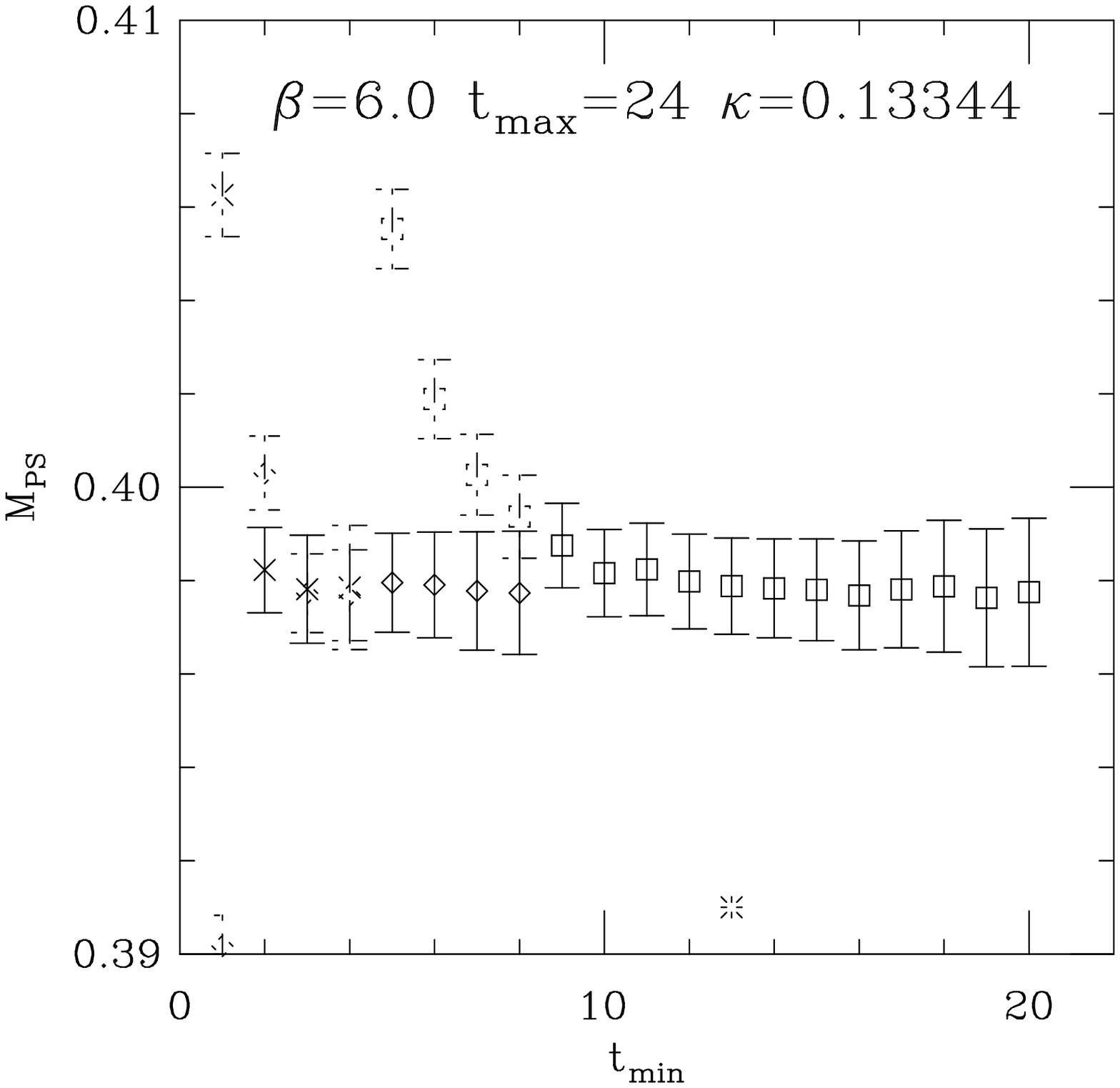}
}
\caption{The ground state masses of the pseudoscalar meson
 as a function of $t_{min}$ for the heaviest mesons at $\beta=5.93$
 and $6.0$. $C_{PA_4}^{FL}$, $C_{PP}^{FL}$, $C_{PP}^{FF}$,
 $C_{PA_4}^{LL}$ and $C_{PP}^{LL}$ were fitted simultaneously, using a
 fit function including ground~($n=1$) and radially excited
 states~($n=2,3$). The solid data points indicate fits for which
 $Q>0.1$. For the dashed results $Q>0.01$ but $<0.1$. The star
 indicates the final value of $t_{min}$ chosen.  We found this fitting
 range to be adequate for all $\kappa$ combinations.  }\label{mslide}
\end{center}
\end{figure}

The ground state masses obtained for the heaviest mesons at the $2$
$\beta$ values are shown in figure~\ref{mslide} as a function of the
initial timeslice for the fit~($t_{min}$).  The results for $M_{PS}^0$
when including only the ground state in the fit function were
consistent with those including radial excitations, and for
simplicity, we used these values in the final analysis shown in
tables~[\ref{mf593}-~\ref{mf60}]. The fitting ranges chosen were
$11-16$ at $\beta=5.93$ and $13-24$ at $\beta=6.0$.  The errors were
generated using $1300$ and $1000$ bootstrap samples for $\beta=5.93$
and $6.0$ respectively. Our results are consistent, to within
$2\sigma$, with the previous determinations of $M_{PS}$ in
reference~\cite{ukqcd,ukqcdd}, which were obtained using a subset of
correlators used here.

\begin{figure}
\begin{center}
\centerline{
\epsfxsize=8.0truecm\epsffile{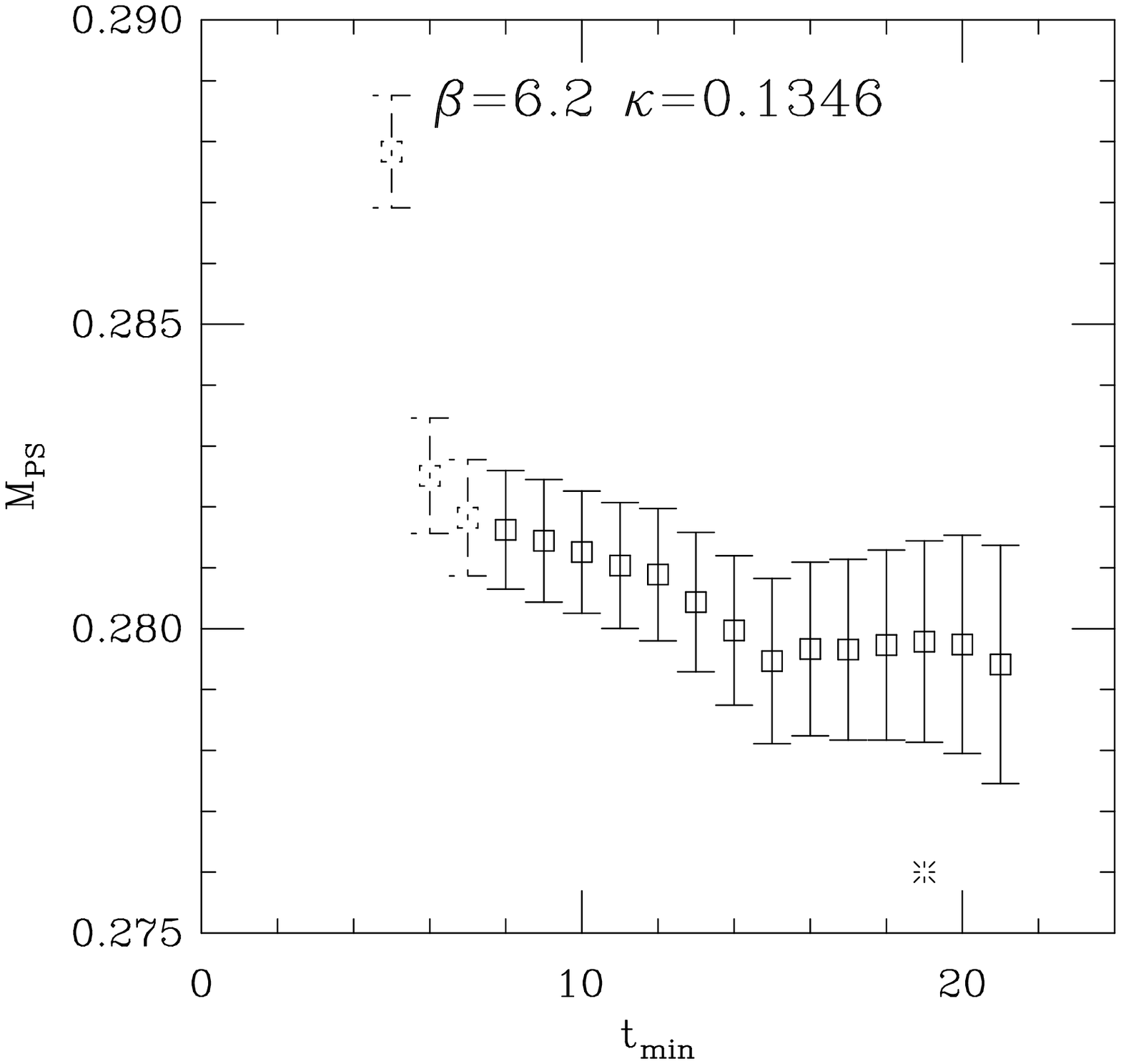}
\epsfxsize=8.0truecm\epsffile{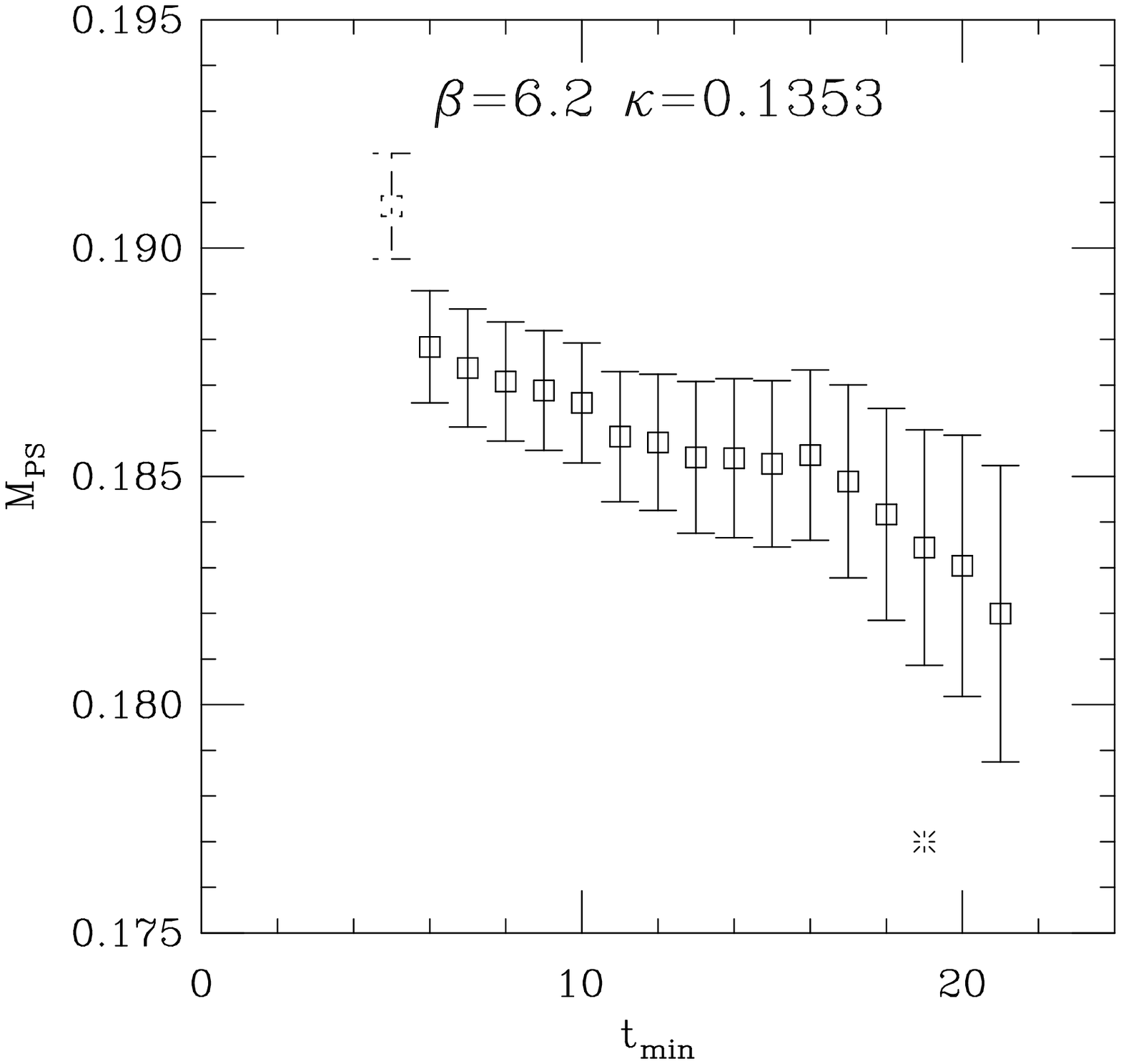}
}
\caption{The ground state masses of the pseudoscalar meson
 as a function of $t_{min}$ for the heaviest and lightest meson at
 $\beta=6.2$.  $C_{PA_4}^{FL}$, $C_{PP}^{FL}$ and $C_{PP}^{FF}$ were
 fitted simultaneously, using a fit function including the ground
 state. The solid data points indicate fits for which $Q>0.1$. For the
 dashed results $Q>0.01$ but $<0.1$. The star indicates the final
 value of $t_{min}$ chosen.  We used this fitting range for all
 $\kappa$ combinations.  }\label{mslide62}
\end{center}
\end{figure}

At $\beta=6.2$, it was not possible to simultaneously fit to all 5
correlators and instead we used $C_{PA_4}^{FL}$, $C_{PP}^{FL}$ and
$C_{PP}^{FF}$ to extract the ground state mass and decay constants and
$C_{PA_4}^{FL}$, $C_{PP}^{FL}$, $C_{PA_4}^{LL}$, $C_{PP}^{LL}$ to
extract the ground state and first excited state contributions to the
correlators being used to extract $c_A$~(see below). In both cases it
was necessary to remove eigenvalues with $c_{cut}=10^{-3}$. For the 3
correlator case only a 1-state fit was successful, as the smearing was
optimised for the ground state and the contributions from excited
states are small. The ground state mass as a function of $t_{min}$ is
shown in figure~\ref{mslide62} for the heaviest and lightest $\kappa$
combinations. For the latter, the mass falls off steadily with
$t_{min}$, even though the $Q$s for the fits are reasonable. Comparing
the fit with the correlators, we find only at $t_{min}=19$ can we be
confident that residual contributions from excited states are below
the statistical errors.  This is true for all $\kappa$, except the
heaviest. We choose $19-25$ for all $\kappa$s. The final values for
$M_P$ are given in table~\ref{mf62}. Fitting to 4 correlators,
including $LL$, $1$, $2$ and $3$ state fits could be performed.

The masses and amplitudes obtained from the 2-state fits can be used
to determine the contribution of the ground and first excited states
to the correlators $C_{PA_4}$ and $C_{PP}$, which are used in the
calculation of $c_A$. The time dependence of this contribution is
interesting to compare with the fitting range chosen to extract
$c_A$. We substitute the parameters from the 2-state fits with the
fitting range $7-15$, $6-24$ and $8-24$ at $\beta=5.93$, $6.0$ and
$6.2$ respectively, into equation~\ref{cap} and calculate the ratio of
this for $n_{max}=1$ and $2$ with the correlators
$C_{PA_4}$~(similarly for $C_{PP}$). The fractional contribution of
the ground state, and the sum of the ground and first excited states,
as a function of timeslice is shown in figures~\ref{fexp1} for
$\beta=5.93$. Note that for a 2-state fit the excited state masses and
amplitudes are likely to contain some contamination from higher
excited states and results in the figures can only give a rough
indication of the fractional contributions.

We see that for the $LL$~($FL$) correlators the ground state dominates
at around timeslice $t_{gs}=12$~($12$), while the first excited state
becomes the dominate radial excitation from approximately timeslice
$t_{1st}=6$ or $7$~($5$).  Repeating this analysis at $\beta=6.0$, we
find $t_{gs}=12$ for $LL$ correlators and $t_{1st}=7$. The smearing at
this $\beta$ was optimised for the ground state and hence it dominates
much earlier, roughly timeslice $8$ for the $FL$ correlators;
$t_{1st}=5$ and the fractional contribution of the first excited state
is very small compared to that for the $LL$.  At $\beta=6.2$, the
smearing was optimised in a similar way and $t_{gs}=10$ for $FL$
compared to $15$ for $LL$ correlators; $t_{1st}\approx8$ and $6/7$ for
$LL$ and $FL$ correlators respectively.  We see that the 1st excited
state dominates at roughly the same physical time; $t_{1st}=6$ at
$\beta=5.93$ corresponds to $t=7$ at $\beta=6.0$ and $\sim9$ at
$\beta=6.2$. We now concentrate on the ground state mass and amplitude
only and drop the superscript $0$ in the following.

The bare improved pseudoscalar decay constant, equation~\ref{fimp},
\begin{eqnarray}
f^{imp} & = & f^{(0)} + c_A af^{(1)}
\end{eqnarray}
can be obtained from the amplitudes and masses:
\begin{eqnarray}
f^{(0)} & = & -2\kappa A^{FL}_{PA_4}\sqrt{\frac{2}{M_P A_{PP}^{FF}}}\\
af^{(1)} & = & \Delta \frac{A_{PP}^{FL}}{A_{PA_4}^{FL}} f^{(0)}
\end{eqnarray}
where 
\begin{equation}
\Delta = \sinh(aM_P)
\end{equation}
or 
\begin{equation}
\Delta = \sinh(aM_P) - \frac{1}{6}(2\sinh(aM_P) -\sinh(a2M_P))
\end{equation}
depending on whether we have used the $O(a^2)$ or $O(a^4)$ definition
of $\partial_4$, respectively, to be consistent with the determination
of $c_A$. The values obtained for $f^{(0)}$ and $af^{(1)}$ extracted
from the 1-state fits are given in
tables~[\ref{mf593}-\ref{mf62}]. For $\beta=5.93$ and $6.2$ the values
of $f^{imp}$ are also given, where $c_A=-0.032(14)$ and $-0.031(5)$
was used, respectively~(obtained by chirally extrapolating $c_A$ - see
next section). The statistical errors in $c_A$ are included by
bootstrapping, into the error estimate of $f^{imp}$.

The bare improved PCAC mass is given by equation~\ref{mpcac},
\begin{eqnarray}
m^{imp}_{PCAC} & = & m^{(0)}_{PCAC} + c_A am^{(1)}_{PCAC}.
\end{eqnarray}
For consistency throughout the analysis we extract $m^{(0)}_{PCAC}$
and $m^{(1)}_{PCAC}$ using the masses and amplitudes extracted in the
1-state fits described above.
\begin{eqnarray}
m^{(0)}_{PCAC} & = & \frac{1}{2}\Delta\frac{A_{PA_4}^{FL}}{A_{PP}^{FL}}\label{m0pcac}\\
a m^{(1)}_{PCAC} & = & \frac{1}{2}\Delta^{(2)}\label{m1pcac}
\end{eqnarray}
where
\begin{eqnarray}
\Delta^{(2)} & = & 2[\cosh(aM_P) - 1]
\end{eqnarray}
and
\begin{eqnarray}
\Delta^{(2)} & = & 2[\cosh(aM_P) - 1] - \frac{1}{6}[\cosh(2aM_P)-4\cosh(aM_P)+3]
\end{eqnarray}
for $O(a^2)$ or $O(a^4)$ temporal derivatives respectively.  The results are
detailed in tables~[\ref{mf593}-\ref{mf62}].  $am_{PCAC}^{imp}$ is
also given for $\beta=5.93$ and $6.0$. The values for $m^{(0)}$ and
$m^{(1)}$ are consistent to within $2\sigma$ with those obtained by
performing a constant fit to $r(t)$ and $s(t)$ directly.

%
\section{Renormalised decay constant and quark mass}
\label{renormfandm}
The renormalised decay constant is obtained from the combination
\begin{eqnarray}
f^{ren} & = & (1+ab_Am_q)Z_Af^{imp}.\label{fren}
\end{eqnarray}
The factor $Z_A$ has been calculated nonperturbatively by the
LANL~\cite{lanl2} and ALPHA~\cite{nonpertza} groups~(as well as by
other groups using axial-Ward identities~\cite{ward} and other
methods~\cite{om}). Their results are consistent to within $2\sigma$
at $\beta=6.0$ and $6.2$. The ALPHA collaboration investigated several
$\beta$ values in the range $6.0$ to $24.0$ and obtained the
interpolating curve,
\begin{eqnarray}
Z_A & = & \frac{1-0.8496g_0^2+0.0610g_0^4}{1-0.7332g_0^2}.
\end{eqnarray}
In perturbation theory $Z_A$ does not depend on $c_A$ to 1-loop.  In
addition, the perturbative value of $Z_A$ is within a few percent of
the nonperturbative value at $\beta\ge6.0$ and hence we do not expect
$Z_A$ to depend significantly on $c_A$ when determined
nonperturbatively over our range of $\beta$s. Thus, we use the
interpolating curve above for $Z_A$. For the errors in $Z_A$, we use
those from the direct simulations of the ALPHA group at $\beta=6.0$
and $6.2$~($Z_A=0.7906(94)$ and $0.807(8)(2)$ respectively), and at
$\beta=5.93$ we assign the same error as at $\beta=6.0$. Although the
interpolating curve is only valid in the range $\beta\ge 6.0$, we do
not expect to incur a significant error by applying it at $\beta=5.93$
since $Z_A$ is not rapidly changing.
%

The coefficient $b_A$ has only been determined nonperturbatively at
$\beta=6.0$ and $6.2$ by the LANL group~\cite{lanl2}. They obtain
$b_A=1.28(3)(4)$ and $1.32(3)(4)$ respectively compared to $1.38(7)$
and $1.34(5)$ from 1-loop tadpole-improved perturbation
theory~\cite{bm,lanl2}, where
\begin{equation}
b_A = \frac{1}{u_0}\left(1+0.8646\alpha\right)
\end{equation}
and we use $\alpha=\alpha_P(1/a)$~\cite{alphap} and the fourth root of
the plaquette for $u_0$. We assign the error in the perturbative
result to be $1\alpha^2_P(1/a)$. The perturbative and nonperturbative
determinations of $b_A$ are consistent and we use the perturbative
results for our 3 $\beta$ values. Finally, we take the quark mass
appearing in equation~\ref{fren} to be $m_{PCAC}^{imp}$.  The results
for $r_0f^{ren}$ at the reference mass $r_0^2M_{PS}^2=3.0$ are given
in table~\ref{chiralf}, where we have applied the same values of $Z_A$
and $b_A$ for results obtained using different values of $c_A$.
 
Values for a renormalised quark mass are normally quoted in terms of
the $\overline{MS}$ scheme at a particular reference scale, we choose
$2$~GeV. We calculate $m^{\overline{MS}}$ in two ways:
nonperturbatively, using the method suggested by the ALPHA
collaboration, where one first calculates the renormalisation-group invariant
quark mass
\begin{equation}
M = Z_M m_{PCAC}^{imp}\label{reninv}
\end{equation}
and then converts to the $\overline{MS}$ scheme at
$2$~GeV~\cite{squark} using (up to) 4-loop perturbation theory. The
ALPHA group have calculated $Z_M$ nonperturbatively~\cite{zm} and
obtained the interpolating curve
\begin{equation}
Z_M = 1.752+0.321(\beta-6)-0.220(\beta-6)^2
\end{equation}
for the range $6.0\le\beta\le 6.5$. The associated ($\beta$ dependent)
uncertainty in $Z_M$ is $1.1\%$. We ignore the additional, $\beta$
independent error of $1.3\%$ as we are only interested in scaling
behaviour and not predictions in the continuum limit. The
determination of $Z_M$ does depend on $c_A$ through the extrapolation
to zero quark mass~(this limit is found using $m_{PCAC}^{imp}$ to
determine $\kappa_c$). The results of the next section will show that
we obtain consistent results for $\kappa_c$ using our values of $c_A$
and those of the ALPHA collaboration and hence we do not expect a
significant error from applying their values for $Z_M$ in our
analysis. The values for $r_0m^{\overline{MS}}(2~GeV)$ for
$r_0^2M_{PS}^2=3.0$ obtained in this way are shown in
table~\ref{chiralf}. The conversion factor from the renormalisation
invariant mass to the $\overline{MS}$ scheme at $2$~GeV is $0.72076$
at 4-loops~\cite{squark}. We apply the same values of $Z_M$ for
results obtained using different values of $c_A$.

We also calculated $m^{\overline{MS}}(2~GeV)$ perturbatively using
\begin{eqnarray}
m^{\overline{MS}}(\mu) & = & \frac{Z_A(1+ab_Am_q)}{Z_P(\mu)(1+ab_Pm_q)}m_{PCAC}^{imp}
\end{eqnarray}
where to 1-loop in tadpole-improved perturbation theory~\cite{zazp,bm,lanl2},
\begin{eqnarray}
Z_A & = & u_0 \left( 1- 0.416 \alpha \right)\\
b_P & = & \frac{1}{u_0}\left(1+0.8763\alpha\right) \\
Z_P(\mu) & = & u_0\left[1+\alpha(\frac{1}{4\pi}\ln(\mu a)^2-1.328)\right]
\end{eqnarray}
We assign errors to the perturbative results of $1\alpha^2_P(1/a)$, as
before.  Using perturbative factors enables a meaningful comparison with
the quark mass determined using $\kappa_c$~(see next section), for
which the renormalisation factor is only known perturbatively. The
resulting values of $m^{\overline{MS}}(2GeV)$ are given in
table~\ref{chiralf}.

\section{Chiral extrapolations.}
\label{secchiralca}
%
%
The improvement scheme we have chosen defines $c_A$ in the limit of zero 
quark mass. We extrapolated our results for $c_A$ at finite quark mass using
the fit function:
\begin{eqnarray}
c_A(m_{q1},m_{q2}) & = & c_0 + c_1(m_{q1}+m_{q2}) +c_2(m_{q1}+m_{q2})^2 + c_3m_{q1}m_{q2},
\label{extrapf}
\end{eqnarray}
where $m_{q1}=M_{PS}^2(\kappa_1)$, the pseudoscalar meson mass for
degenerate quarks with $\kappa=\kappa_1$.  We performed correlated
fits for all $t_{ex}$, $t_{gs}$ combinations which gave reasonable $Q$
values at finite quark mass. Higher order terms in the fit function
were included successively, starting with a constant fit. Each fit
function was applied to the set of $c_A(m_{q1},m_{q2})$ starting with
the lightest $m_{q1}$ and $m_{q2}$. 

The results are detailed in
tables~\ref{chiralttwo},~\ref{chiraltthree} and~\ref{chiraltfour} for
$\beta=5.93$, $6.0$ and $6.2$, respectively, where for each $t_{ex}$
and $t_{gs}$ combination the fits which cover the largest quark mass
range are given. In the case of a constant fit, $c_A(\kappa_c)$ is an
average of the values at finite quark mass; we prefer to quote the
value for the heaviest $\kappa$ value, which is consistent but has a
larger error.

At $\beta=5.93$, we encountered difficulties performing the chiral
extrapolations. Often the fit would lie above or below all the data
points. In most cases, the problem was solved by fitting to only the
degenerate $\kappa$ combinations, and hence reducing the correlations
between data points. If this was not sufficient we also dropped the
smallest eigenvalues from the covariance matrix, as described in the
previous section.

In some cases after eigenvalues were dropped there were not enough
degrees of freedom remaining for the fit to be performed.  Instead we
performed a linear and quadratic ($c_3=0$) uncorrelated fit to the
full quark mass range. These results are also given in the tables. For
the comparison of $c_A(\kappa_c)$ as a function of $t_{ex}$~(see
figure~\ref{figc}), we take the result from the linear uncorrelated
fit, unless there is a statistically significant difference between
the two fits, in which case we use the quadratic fit. We also give in
the tables the results of uncorrelated fits for comparison in the
cases where we could not fit over the full quark mass range using a
correlated fit. The final value for $c_A$ at each $\beta$ is chosen
from the set of successful correlated fits only.

Compared to the statistical errors in $c_A$ at finite $\kappa$,
fitting to only a limited number of data points gave rise to large
errors in the chiral limit.  At $\beta=6.0$ and $6.2$, we followed the same
procedure. The correlations do not seem to be so severe, however, and
it was possible to fit to the non-degenerate and degenerate $\kappa$
combinations and no eigenvalues needed to be set to zero.

%
%
Using $c_A(\kappa_c)$, we obtain $m_{PCAC}^{imp}$ and perform a
consistency check by chirally extrapolating $r_0^2M_{PS}^2$ as a
function of $r_0m_{PCAC}^{imp}$; from the definition of
equation~\ref{mpcac}, and equations~\ref{m0pcac} and~\ref{m1pcac},
$M_{PS}$ and $m_{PCAC}$ should vanish at the same point. In the first
set of fits we used the same functional form as in
equation~\ref{extrapf} with $m_{q1}=m_{PCAC}^{imp}(\kappa_1)$, and
used the same procedure as above. We repeated the analysis using $c_A$
as determined by the ALPHA collaboration.  The results are given in
table~\ref{kc} for $\beta=5.93$ and~\ref{kc2} for $\beta=6.0$ and
$6.2$.

\begin{figure}
\begin{center}
\centerline{
\epsfxsize=8.0truecm\epsffile{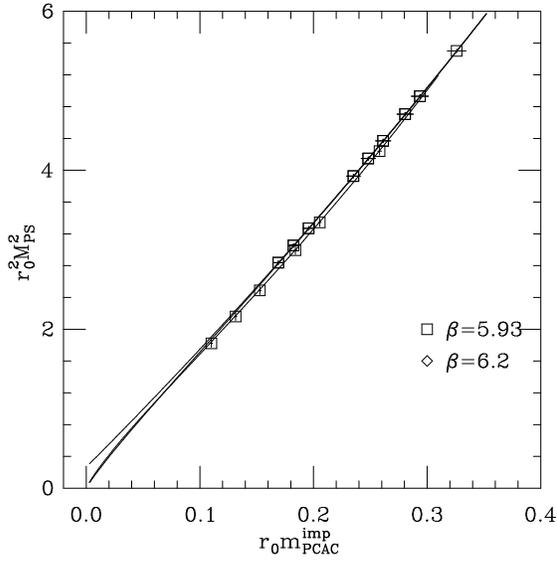}
\epsfxsize=8.0truecm\epsffile{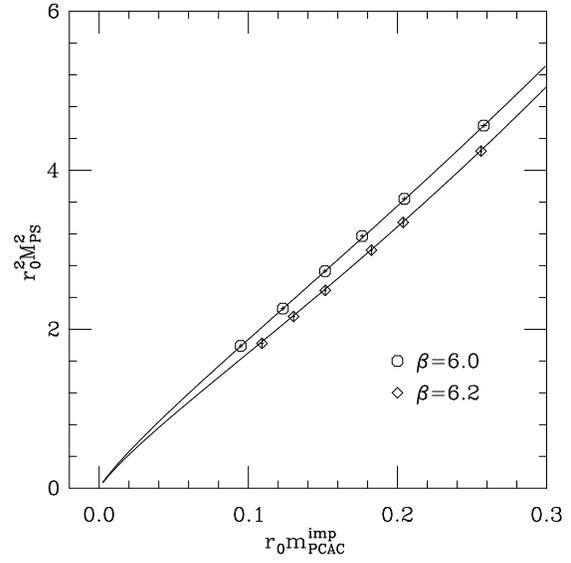}
}
\centerline{(a)\hspace{10cm}(b)}
\centerline{
\epsfxsize=8.0truecm\epsffile{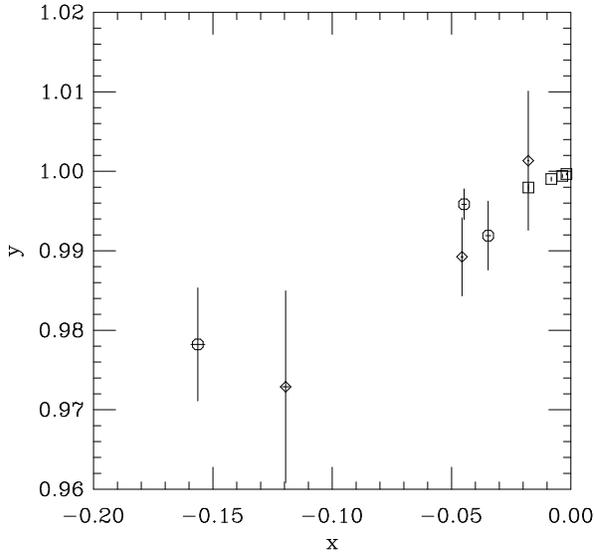}
\epsfxsize=8.0truecm\epsffile{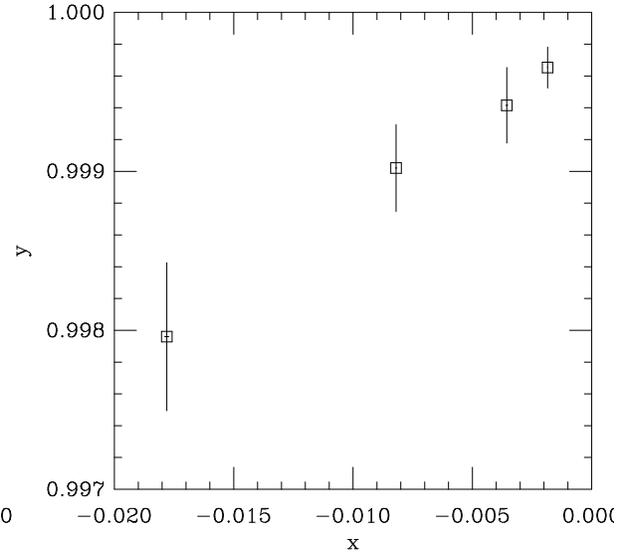}
}
\centerline{(c)\hspace{10cm}(d)}
\caption{$r_0^2M_{PS}^2$ vs $r_0m_{PCAC}^{imp}$. (a) using $c_A$ as determined in this paper and (b) as determined by the ALPHA collaboration.(c) shows the quantity $y$
as a function of $x$ as defined in equations~\protect\ref{ry}
and~\protect\ref{rx} for all 3 $\beta$ values. (d) presents the
results for $\beta=5.93$ only. Note that in (c) and (d)
$m_{PCAC}^{(0)}$ has been used.}\label{fig2c}
\end{center}
\end{figure}

At $\beta=5.93$ we see that whether using a linear or quadratic fit
function~(with successively wider ranges in quark mass), there is a
non-zero value for $c_0$. Fits with $c_0$ forced to zero were
unsuccessful. This effect was also seen at $\beta=6.0$ and $6.2$.  We
therefore tried to resolve the presence of quenched chiral logarithms.
Following the analysis of the CP-PACS collaboration~\cite{cppacs}
we computed the quantities
\begin{eqnarray}
y & = & \frac{2m_{q1}}{(m_{q1}+m_{q2})} \frac{M_{P,12}^2}{M_{P,11}^2} \times
\frac{2m_{q1}}{m_{q1}+m_{q2}}\frac{M_{P,12}^2}{M_{P,22}^2} \label{ry}\\
x & = & 2 + \frac{m_{q1}+m_{q2}}{m_{q1}-m_{q2}}\ln(\frac{m_{q2}}{m_{q1}})
\label{rx}
\end{eqnarray}
which are related by $y=1+\delta\cdot x + O(m^2)$. Any significant
deviation of $y$ from $1$ indicates a non-zero value for $\delta$, the
chiral log term. Figure~\ref{fig2c} shows that we observe a clear
non-zero slope with $\delta$ very roughly in the range $0.1-0.2$, with
$y$ deviating from $1$ by more than $3\sigma$s at
$\beta=5.93$. Previous estimates of $\delta$ vary in the range of
$0.08{-}0.12$ from CP-PACS~\cite{cppacs}, $\sim 0.06$ from Bernard
et. al.~\cite{bernard} and $0.065(13)$ from Bardeen
et. al.~\cite{bardeen}. This analysis motivates us to include a log
term in the fitting function in order to set $c_0=0.$:
\begin{eqnarray}
M_P^2 & = & c_1(m_{q1}+m_{q2}) - c_{log}(m_{q1}+m_{q2})[\ln(2m_{q1})+\frac{m_{q2}}{m_{q2}-m_{q1}}\ln(\frac{m_{q1}}{m_{q2}})] \nonumber\\ & & + c_2(m_{q1}+m_{q2})^2 + c_3m_{q1}m_{q2}\label{chirall}
\end{eqnarray}
The results of these fits are also shown in tables~\ref{chirals}
and~\ref{kc2}. 
We present the best fits in figures~\ref{fig2c} using both our values
for $c_A$ and those of the ALPHA collaboration. In figure~(a) a
quadratic fit (function 3 to the 8 lightest data points) to the
$\beta=5.93$ data set is included to illustrate the non-zero intercept
found when there are no chiral logs terms in the fit.  Since
$m_{PCAC}^{imp}$ is a bare mass there is no significance in the fact
that there is better agreement between the results at different
$\beta$ values when our values for $c_A$ are used compared to those of
the ALPHA collaboration.

We also extrapolated $m_{PCAC}^{imp}$ and $M_{PS}^2$ as function of
$1/\kappa$ in order to extract $\kappa_c$. The latter can be used to
obtain $m^{\overline{MS}}$ independent of $c_A$. There are no chiral
logarithms expected for the PCAC mass and we use
equation~\ref{extrapf} for the extrapolations with $c_0=0$ and
$m_q=\tilde{m}_q$ defined as
\begin{eqnarray}
\tilde{m}_q & = & m'_q(1+b_m m'_q)\label{mtilde}\\
m'_q & = & \frac{1}{2}\left(\frac{1}{\kappa} - \frac{1}{\kappa_c}\right)
\end{eqnarray}
where $\kappa_c$ is an additional parameter in the fit. $b_m$ is known to
1-loop in perturbation theory~\cite{bm}. Including tadpole-improvement
\begin{eqnarray}
b_m & = & -0.5 - 0.686\alpha.
\end{eqnarray}
We obtain $b_m=-0.730$, $-0.761$ and $-0.776$ at $\beta=6.2$, 6.0 and
5.93, respectively using $\alpha_P(1/a)$. The results of the fits to
$m_{PCAC}^{imp}$ are shown in table~\ref{kc}. To extract $\kappa_c$
from $M_P^2$ we use equation~\ref{chirall} with
equation~\ref{mtilde}. The results of the fits are also given in
table~\ref{kc}. For comparison we present the results of fits without
the chiral logs.

We see a significant drop in the value of $\kappa_c$ at $\beta=5.93$
and $6.0$ from $M_{PS}^2$ when log terms are included. For $\beta=6.2$
the results are consistent to within $2\sigma$. The values of
$\kappa_c$ extracted from the PCAC mass and $M_P^2$ should be
consistent. We see that this is the case when comparing the PCAC mass
results with those of the chiral log fits to the pseudoscalar meson
mass.  Disagreement between the two determinations of $\kappa_c$ when
log terms are omitted for $M_{PS}^2$ has been noted and discussed
previously in reference~\cite{lanl2} and also in
reference~\cite{cppacs}.

$\kappa_c$ has been extracted previously by UKQCD from the simulations
at $\beta=6.0$ and $6.2$ in reference~\cite{ukqcd} and $\beta=5.93$ in
reference~\cite{ukqcdd}. In those works uncorrelated linear
extrapolations~(without chiral logs) were performed to $M_{PS}^2$
using all $\kappa$ combinations and with $b_m$ calculated using
boosted perturbation theory~(the dependence of $\kappa_c$ on the value
of $b_m$ used was investigated and found to be
small). $\kappa_c=0.135202(11)$, $0.135252^{+16}_{-9}$ and
$0.135815^{+17}_{-14}$ was obtained at $\beta=5.93$, $6.0$ and $6.2$,
respectively. These values are consistent to within $2\sigma$ with our
results without chiral logs.

%

The ALPHA collaboration has determined $\kappa_c$ from
$m_{PCAC}^{imp}$ at $\beta=6.0$ and $6.2$~\cite{alpha}: they find
$\kappa_c=0.135196(14)$ and $0.135795(13)$ respectively. These results
are consistent, within $2\sigma$, with those in table~\ref{kc}
obtained using the ALPHA value for $c_A$.

\section{Renormalised quark mass from $\kappa_c$}
From the extrapolations of $r_0^2M_{PS}^2$ versus
$r_0\tilde{m}_q$~(equation~\ref{mtilde}) we can extract the value of
$\tilde{m}_q$ which corresponds to $r_0^2M_{PS}^2=3.0$. This 
can be converted to a value for $m^{\overline{MS}}$ using
\begin{eqnarray}
m^{\overline{MS}}(\mu) & = & Z_m^{\overline{MS}}(a\mu)\tilde{m}.
\end{eqnarray}
The renormalisation factor is only known 1-loop in perturbation
theory~\cite{zmpert}.  Using tadpole-improvement,
\begin{eqnarray}
Z_m(1/a) & = & \frac{1}{u_0}\left[ 1+1.002\alpha\right].
\end{eqnarray}
The results for $m^{\overline{MS}}(2GeV)$ are given in table~\ref{chiralf}.
\end{appendix}



\begin{table}
\caption{Simulation details.}\label{sim}
\begin{tabular}{ccccccc}
$\beta$ & Volume & $n_{confs}$ & $C_{SW}$ & smearing & $\kappa_l$ & $L$~(fm) \\\hline
5.93 & $16^3\times 32$ & 684 & 1.82  & $LL$, $FL$, $FF$ & 0.1327, 0.1332, 0.1334, 0.1337,0.1339 & 1.7 \\
6.0 & $16^3\times 48$ & 496 & 1.77 & $LL$, $FL$, $FF$ & 0.13344, 0.13417, 0.13455 & 1.5 \\
 & $32^3\times 64$ & 70  &  & $SL$, $SS$ &  & 3.0 \\
6.2 & $24^3\times 48$ & 214 & 1.61 & $LL$, $FL$, $FF$ & 0.1346, 0.1371, 0.13745 & 1.6
\end{tabular}
\end{table}

\begin{table}
\caption{Details of the calculation of $c_A$ at $\beta=5.93$ 
for the heaviest $\kappa$ value, $\kappa=0.1327$ and $FL$ and
$LL$ correlators. Standard, $O(a^2)$ derivatives are used for the
temporal derivatives.}\label{tone}
\begin{tabular}{ccccc|cc}
\multicolumn{7}{c}{$\partial=O(a^2)$}\\\hline\hline
$t_{ex}$ & $t_{gs}$ & $s(t_{ex}){-}s(t_{gs})$ & 
$r(t_{ex}){-}r(t_{gs})$ & $c_A$ & $c_A$ LANL & Q \\
& & & & $-\frac{\Delta r}{\Delta s}$ & fit $t_{ex}$ to
$t_{gs}$ & \\\hline\hline
\multicolumn{7}{c}{$FL$}\\\hline
3  & 12 & 0.552(6) & 0.0770(11) & -0.139(2) & -0.131(2) & 0. \\
4  &  &  0.348(5) & 0.0354(9) & -0.102(2) & -0.101(3) & 0. \\
5  &  &  0.181(4) & 0.0159(6) & -0.088(3) & -0.089(3) & 0.02 \\
6  &  &  0.092(4) & 0.0069(6) & -0.074(7) & -0.078(6) & 0.04 \\
7  &  &  0.052(3) & 0.0027(6) & -0.051(11) & -0.058(12) & 0.08 \\
8  &  &  0.023(3) & 0.0012(5) & -0.052(24) & -0.065(33) & 0.09 \\
9  &  &  0.014(3) & 0.0014(5) & -0.104(48) & -0.134(55) & 0.30 \\\hline
3  & 14 & 0.552(6) & 0.078(1) & -0.141(2) & -0.130(2) & 0. \\
4  &  & 0.348(5) & 0.0365(9) & -0.105(2) & -0.101(3) & 0. \\
5  &  & 0.181(4) & 0.0170(7) & -0.094(4) & -0.090(3) & 0.04 \\
6  &  & 0.092(3) & 0.0079(6) & -0.086(7) & -0.081(6) & 0.05 \\
7  &  & 0.052(3) & 0.0037(6) & -0.073(11) & -0.068(11) & 0.05 \\
8  &  &  0.022(3) & 0.0023(6) & -0.101(28) & -0.091(33) & 0.05\\
9  &  &  0.013(3) & 0.0025(6) & -0.188(67) & -0.162(57) & 0.39\\\hline
\multicolumn{7}{c}{$LL$}\\\hline
3  & 12 & 1.78(1) & 0.437(3) & -0.246(1) & -0.267(3) & 0.\\
4  &  & 0.758(8) & 0.134(2) & -0.177(1) &  -0.177(3)& 0.\\
5  &  &  0.314(5) & 0.0416(8) & -0.132(2) & -0.133(3)& 0.\\
6  &  &  0.142(4) & 0.0141(7) & -0.100(5) & -0.105(5)& 0.\\
7  &  &  0.073(4) & 0.0048(7) & -0.066(9) & -0.073(8) & 0.09\\
8 &   &  0.032(3) & 0.0018(6) & -0.058(17) & -0.067(20) & 0.05\\
9 &   &  0.018(3) & 0.0018(6) & -0.102(36) & -0.125(39) & 0.32\\\hline
3  & 14 & 1.78(1) & 0.438(3) & -0.246(1) &  -0.266(3) & 0.  \\
4  &  & 0.757(8) & 0.135(2) & -0.178(1) & -0.176(3) & 0. \\
5  &  &  0.314(5) & 0.0427(9) & -0.136(3) & -0.132(3) & 0. \\
6  &  &  0.141(4) & 0.0152(7) & -0.108(5) & -0.106(4) & 0.0  \\
7  &  &  0.072(4) & 0.0059(6) & -0.082(9) & -0.079(8) & 0.07 \\
8  &  &  0.031(4) & 0.0030(6) & -0.096(20) & -0.082(19) & 0.04 \\
9  &  &  0.017(3) & 0.0029(6) & -0.174(49) & -0.145(40) & 0.33 
\end{tabular}
\end{table}

\begin{table}
\caption{The same as for table~\protect\ref{tone} using improved~($O(a^4)$) lattice derivatives. }\label{ttwo}
\begin{tabular}{ccccc|cc}
\multicolumn{7}{c}{$O(a^4)$}\\\hline\hline
$t_{ex}$ & $t_{gs}$ & $s(t_{ex}){-}s(t_{gs})$ &
$r(t_{ex}){-}r(t_{gs})$ & $c_A$ & $c_A$ LANL & Q \\ & & & &
$-\frac{\Delta r}{\Delta s}$ & fit $t_{ex}$ to $t_{gs}$ &
\\\hline\hline
\multicolumn{7}{c}{$FL$}\\\hline
4  & 12 & 0.296(5) & 0.0103(10) & -0.035(3) & -0.036(3) & 0.05 \\
5  &  &  0.149(4) & 0.0048(7) & -0.032(5) & -0.033(4) &0.04 \\
6  &  & 0.077(4) & 0.0017(7) & -0.022(9) & -0.024(9) & 0.04 \\
7  &  & 0.045(4) & -0.00002(71) & +0.0004(16) & -0.012(16) & 0.03 \\
8  &  & 0.018(3) & -0.00009(61) & +0.005(35) & -0.035(817) & 0.02 \\\hline
4  & 14 & 0.295(5) & 0.0116(10) & -0.039(3) & -0.037(2)& 0.07 \\
5  &  &  0.148(4) & 0.0061(8) & -0.041(5) & -0.035(4) & 0.05 \\
6  &  &  0.076(3) & 0.0030(7) & -0.040(10) & -0.029(9) & 0.04 \\
7  &  &  0.045(4) & 0.0013(6) & -0.030(15) & -0.025(15) & 0.02 \\
8 &   &  0.017(3) & 0.0012(6) & -0.072(45) & -0.066(264) & 0.03\\\hline
\multicolumn{7}{c}{$LL$}\\\hline
4  & 12 & 0.483(6) & -0.0276(12) & +0.057(3) & +0.034(2) & 0. \\
5  &  &  0.227(5) & 0.0010(8) & -0.004(4) & -0.007(3) & 0.01  \\
6  &  &   0.112(4) & 0.0020(8) & -0.018(7) & -0.020(6) & 0.03\\
7  &  &  0.062(4) & 0.0001(8) & -0.002(13) & -0.012(12) & 0.02 \\
8 &    &  0.025(4) & -0.0001(6) & +0.003(26) & -0.026(33) & 0.02 \\\hline
4  & 14 & 0.481(6) & -0.0262(12) & +0.054(3) & +0.032(2) & 0. \\
5  &  & 0.225(5) & 0.0024(9) & -0.011(4) & -0.010(3) & 0.0  \\
6  &  & 0.110(4) & 0.0034(8) & -0.031(7) & -0.024(6) & 0.03 \\
7  &  & 0.060(4) & 0.0015(7) & -0.025(12) & -0.022(11) & 0.02\\
8  &  &  0.024(4) & 0.0013(7) & 0.056(31) & -0.046(30) & 0.01
\end{tabular}
\end{table}

\begin{table}
\caption{$c_A$ in the chiral limit for $\beta=5.93$ using both $O(a^2)$ and $O(a^4)$ derivatives. Columns $1{-}7$ give details
of correlated chiral extrapolations of $c_A$ determined from mesons
with degenerate quark masses. Included are the function used~(in terms
of the index of the highest order coefficient included - see
equation~\protect\ref{extrapf}), the number of eigenvalues dropped
from the covariance matrix, $n_{drop}$ and the number of data
points used in the fit, $n_\kappa$. In some cases a linear and
quadratic uncorrelated fit~(function 1. and 2., respectively) was
performed. The results for $c_A(\kappa_c)$ for the uncorrelated fits
are given in columns $8{-}9$.}\label{chiralttwo}
\begin{tabular}{ccccccc|cc}
\multicolumn{9}{c}{$O(a^2)$}\\\hline\hline
$t_{ex}$ & $t_{gs}$ & func. & $n_{\kappa}$ & $c_A(\kappa_c)$ & Q & $n_{drop}$ & uncor func 1. & uncor func 2. \\\hline\hline
\multicolumn{9}{c}{$FL$}\\\hline
5  & 12 &  &  &  &  &  & -0.068(6) & -0.079(10)\\
6  &    & 1. & 4 & -0.048(10) & 0.72 & 0 & -0.046(15) & -0.037(27)\\
7  &    & 2. & 4 & -0.002(29) & 0.78 & 0 & +0.006(30) & +0.036(56)\\\hline
5  & 14 &  &  &  &  &  & -0.069(6) & -0.079(10)\\
6  &    & 2. & 4 & -0.045(21) & 0.68 & 0 & -0.049(15) & -0.040(28)\\
7  &    & 2. & 4 & +0.023(49) & 0.73 & 0 & -0.006(31) & +0.033(61)\\\hline
\multicolumn{9}{c}{$LL$}\\\hline
7  & 12 & 2. & 4 & -0.006(30) & 0.74 & 0 & -0.019(22) & +0.002(40)\\
8  &    & 1. & 4 & +0.011(50) & 0.98 & 0 & +0.016(61) & +0.046(149)\\\hline
7  & 14 & 2. & 4 & -0.030(23) & 0.45 & 0 & -0.029(21) & +0.000(39)\\
8  &    & 1. & 4 & +0.001(64) & 0.98 & 0 & -0.002(68) & +0.068(295)\\\hline\hline
\multicolumn{9}{c}{$O(a^4)$}\\\hline\hline
\multicolumn{9}{c}{$FL$}\\\hline
4  & 12 & 1. & 5 & -0.050(3) & 0.05 & 0 &  & \\
5  &    &   &  &          &     &  & -0.051(8) & -0.063(13)\\
6  &    &   &  &          &     &  & -0.018(21)& +0.000(38)\\\hline
4  & 14 & 1. & 5 & -0.050(3) & 0.2  & 1 &  & \\
5  &    &   &  &          &     &  & -0.052(8) & -0.064(13)\\
6  &    & 2. & 5 & -0.022(19)& 0.78 & 0 & & \\\hline
\multicolumn{9}{c}{$LL$}\\\hline
5  & 12 & 1. & 4 & -0.039(4) & 0.01 & 0 & -0.041(6) & -0.047(10)\\
6  &    & 2. & 5 & -0.028(13)& 0.85 & 0 & &\\\hline
6  & 14 & 2. & 5 & -0.032(14)& 0.70 & 0 &  &
\end{tabular}
\end{table}


\begin{table}
\caption{$c_A$ in the chiral limit for $\beta=6.0$. The details are the
same as in table~\protect\ref{chiralttwo}. No eigenvalues were dropped
from the covariance matrix. In the case of a constant fit the (*)
indicate the value of $c_A$ from the heaviest $\kappa$ value is taken
rather than the value from the fit.}\label{chiraltthree}
\begin{tabular}{cccccc|cc}
\multicolumn{8}{c}{$O(a^4)$}\\\hline\hline
$t_{ex}$ & $t_{gs}$ & form & $n_{\kappa}$ & $c_A(\kappa_c)$ & Q & u
lin & u quad \\\hline\hline
\multicolumn{8}{c}{$FL$}\\\hline
4  & 12 & 1. & 4 & -0.124(58) & 0.94 & -0.095(52) & -0.154(120)\\
4  & 16 & 1. & 4 & -0.145(59) & 0.95 & -0.113(54) & -0.188(153)\\\hline
\multicolumn{8}{c}{$LL$}\\\hline
6  & 12 & 2. & 4 & -0.084(53) & 0.69 & -0.056(17) & -0.069(33)\\
7 &    & 1. & 5 & -0.106(48) & 0.97 & -0.114(52) & -0.166(172)\\
6  & 16 & 1. & 4 & -0.065(12) & 0.99 & -0.063(16) & -0.067(30)\\
7 &    & 3. & 6 & -0.147(66) & 0.91 &  & \\\hline
\multicolumn{8}{c}{$SL$}\\\hline
4 & 12 & 1. & 6 & +0.002(12) & 0.1 &   &  \\
5 &    & 0 & 6 & +0.008(16) & 0.17&  & \\
  &    &   &   & +$0.006(18)^*$ &     &    &  \\
4 & 16 & 3. & 6 & -0.008(19) & 0.92 &  & \\
5 &    & 1. & 5 & +0.017(25) & 0.74 & +0.014(40) & +0.101(77)\\\hline
\end{tabular}
\end{table}

\begin{table}
\caption{$c_A$ in the chiral limit for $\beta=6.2$. The details are the
same as in table~\protect\ref{chiralttwo}. No eigenvalues were dropped
from the covariance matrix. In the case of a constant fit the (*)
indicate the value of $c_A$ from the heaviest $\kappa$ value is taken
rather than the value from the fit. }\label{chiraltfour}
\begin{tabular}{cccccc}
\multicolumn{6}{c}{$O(a^4)$}\\\hline\hline
$t_{ex}$ & $t_{gs}$ & form & $n_{\kappa}$ & $c_A(\kappa_c)$ & q \\\hline\hline
\multicolumn{6}{c}{$FL$}\\\hline
4  & 16 & 1. & 6 & -0.021(11) & 0.65 \\
5  &    & 1. & 6 & -0.048(35) & 1.0 \\\hline
\multicolumn{6}{c}{$LL$}\\\hline
6  & 16 & 3. & 6 & -0.043(7) & 0.75 \\
7  &    & 0. & 6 & -0.029(4) & 0.76 \\
   &    &  &  &  $-0.031(5)^*$ &  \\
8 &  & 0 & 6 & -0.028(7) & 0.89 \\
   &    &  &  &  $-0.029(9)^*$ &  \\\hline
\end{tabular}
\end{table}

%
\begin{table}
\caption{The pseudoscalar meson masses, decay 
constants and $am_{PCAC}$ extracted at $\beta=5.93$. $a^2m_{PCAC}^{(1)}$
and $a^2f^{(1)}$ were extracted using $O(a^4)$ derivatives. The error in $c_A$
is included in the results for $am_{PCAC}^{imp}$ and $af^{imp}$.\label{mf593}}
\begin{tabular}{ccccccccc}
 $\kappa_1$ & $\kappa_2$ &   $aM_{PS}$ & $am_{PCAC}^{(0)}$ & $a^2m_{PCAC}^{(1)}$ & $am_{PCAC}^{imp}$ & $af^{(0)}$ & $a^2f^{(1)}$ & $af^{imp}$\\\hline
0.1327 & 0.1327 & 0.4948(10) & 0.0727(1) & 0.1223(5)  & 0.0688(18) & 0.131(1) & 0.219(2) & 0.124(3)\\
0.1332 & 0.1327 & 0.4684(10) & 0.0655(1) & 0.1096(5) & 0.0619(16) & 0.128(1) & 0.214(2) & 0.121(3)\\
0.1334 & 0.1327 & 0.4575(10) & 0.0626(1) & 0.1046(5) & 0.0592(15) & 0.127(1)& 0.212(2) & 0.120(3)\\
0.1332 & 0.1332 & 0.4409(11) & 0.0583(1) & 0.0972(5) & 0.0551(14) & 0.126(1)& 0.209(2) & 0.119(3)\\
0.1334 & 0.1332 & 0.4296(11) & 0.0553(1) & 0.0922(5) & 0.0524(13) & 0.125(1)& 0.207(2) & 0.118(3)\\
0.1334 & 0.1334 & 0.4179(12) & 0.0524(1) & 0.0873(5) & 0.0496(13) & 0.123(1)& 0.205(2) & 0.117(3)\\
0.1337 & 0.1337 & 0.3814(13) & 0.0436(1) & 0.0727(5) & 0.0413(11) & 0.120(1)& 0.200(2) & 0.114(3)\\
0.1339 & 0.1337 & 0.3686(13) & 0.0407(1) & 0.0679(5) & 0.0385(10) & 0.119(1)& 0.198(2) & 0.112(3)\\
0.1339 & 0.1339 & 0.3553(14) & 0.0377(1) & 0.0631(5) & 0.0357(9)  & 0.118(2)& 0.197(2) & 0.111(3)
\end{tabular}
\end{table}

\begin{table}
\caption{Same as in table~\protect\ref{mf593} for $\beta=6.0$. \label{mf60}}
\begin{tabular}{ccccccc}
 $\kappa_1$ & $\kappa_2$  & $aM_{PS}$ &  $am_{PCAC}^{(0)}$ & $a^2m_{PCAC}^{(1)}$ &$af^{(0)}$ & $a^2f^{(1)}$\\\hline
0.13344 & 0.13344 & 0.3979(10) & 0.0532(4) & 0.0791(4) & 0.111(1) & 0.164(2)\\
0.13417 & 0.13344 & 0.3555(12) & 0.0425(4) & 0.0632(4) & 0.107(1) & 0.159(2)\\
0.13455 & 0.13344 & 0.3317(14) & 0.0366(3) & 0.0550(4) & 0.105(1) & 0.157(2)\\
0.13417 & 0.13417 & 0.3078(13) & 0.0317(3) & 0.0474(4) & 0.103(1) & 0.153(2)\\
0.13455 & 0.13417 & 0.2801(15) & 0.0258(2) & 0.0392(4) & 0.100(1) & 0.152(2)\\
0.13455 & 0.13455 & 0.2493(18) & 0.0201(2) & 0.0311(4) & 0.098(1)& 0.151(3)
\end{tabular}
\end{table}

\begin{table}
\caption{The same as in table~\protect\ref{mf593} for $\beta=6.2$. \label{mf62}}
\begin{tabular}{ccccccccc}
 $\kappa_1$ & $\kappa_2$  & $aM_{PS}$ &  $am_{PCAC}^{(0)}$ & $a^2m_{PCAC}^{(1)}$ & $am_{PCAC}^{imp}$ &$af^{(0)}$ & $a^2f^{(1)}$ & $af^{imp}$\\\hline
0.1346 & 0.1346 & 0.2798(17) & 0.0363(1) & 0.0391(5) & 0.0351(2)& 0.079(1) & 0.085(2) & 0.076(1)  \\
0.1351 & 0.1346 & 0.2484(19) & 0.0289(1) & 0.0309(5) & 0.0279(2) & 0.076(1) & 0.081(2) & 0.073(1)  \\
0.1353 & 0.1346 & 0.2351(21) & 0.0259(1) & 0.0276(5) & 0.0250(2) & 0.074(1) & 0.079(2) & 0.072(1) \\
0.1351 & 0.1351 & 0.2144(22) & 0.0215(1) & 0.0230(5) & 0.0208(2) & 0.073(1) & 0.078(2) & 0.070(1)\\
0.1353 & 0.1351 & 0.1997(23) & 0.0185(1) & 0.0199(5) & 0.0178(1) & 0.071(1) & 0.077(2)& 0.069(1) \\
0.1353 & 0.1353 & 0.1834(26) & 0.0155(1) & 0.0168(5) & 0.0150(1) & 0.070(1) & 0.076(3)& 0.068(1)
\end{tabular}
\end{table}

\begin{table}
\caption{Parameters from $r_0^2M_{PS}^2$ vs $r_0m_{PCAC}^{imp}$ at 
$\beta=5.93$. The function L2 etc refers to the fit function in
equation~\protect\ref{chirall}, $n_\kappa$ to the number of data
points in the fit and $n_{drop}$ to the number of eigenvalues omitted
from the covariance matrix. Where the errors on the coefficients are
not symmetric, assymetric errors are quoted.\label{chirals}}
\begin{tabular}{ccccccccc}
fit & $n_{\kappa}$ & $Q$ & $n_{drop}$ & $c_0$ & $c_1$ & $c_{log}$ &
$c_2$ & $c_3$\\\hline
\multicolumn{9}{c}{$c_A=NP$}\\\hline
1 & 3 & 0.19 & 0 & 0.08(3) & 8.2(2) &   &   &\\
2 & 5 & 0.24 & 0 & 0.23($^4_3$) & 7.3(2) &   & 1.2(1) & \\
3 & 8 & 0.83 & 2 & 0.27(3) & 7.1(2) &   & 1.6(2) & -1.1($^3_4$)\\\hline
L2 & 5 & 0.14 & 0 &        & 7.5(2) & 1.0(2) & 2.3(3) &   \\
   & 9 & 0.20 & 2 &        & 7.5(1) & 1.0(1) & 2.4($^3_2$) &    \\\hline
\multicolumn{9}{c}{$c_A=ALPHA$}\\\hline
1 & 3 & 0.19 & 0 & 0.17(3) & 8.9(6) &    &   & \\
2 & 5 & 0.27 & 0 & 0.24(4) & 8.4(1) &    & 0.8(1) & \\
3 & 8 & 0.59 & 2 & 0.27($^4_3$) & 8.3(1) &    & 0.69(13) & 1.2($^3_4$) \\\hline
L2 & 4 & 0.20 & 0 &     & 8.2(2) & 1.6(2) & 3.2($^4_5$) &   \\
L3 & 7 & 0.27 & 1 &     & 8.4(1) & 1.3(2) & 1.8(2) & 3.4(5) \\
\end{tabular}
\end{table}

\begin{table}
\caption{Parameters from $r_0^2M_{PS}^2$ vs $r_0m_{PCAC}^{imp}$ at $\beta=6.0$ and $6.2$. This time only for the best fits. The details are the same
as in table~\protect\ref{chirals}.\label{kc2}}
\begin{tabular}{ccccccccc}
fit & $n_{\kappa}$ & $Q$ & $n_{drop}$ & $c_0$ & $c_1$ & $c_{log}$ &
$c_2$ & $c_3$\\\hline
\multicolumn{9}{c}{$\beta=6.0$, $c_A=ALPHA$}\\\hline
3 & 6 & 0.76 & 0 & 0.29(7) & 7.7(4) &   & 1.9(1.1) & -3.2($^{2.3}_{2.2}$)\\
L3& 6 & 0.39 & 0 &        & 7.7(4) & 1.7(4) & 3.7($^{1.5}_{1.6}$) & -1.5(2.0) \\\hline
\multicolumn{9}{c}{$\beta=6.2$, $c_A=NP$}\\\hline
3 & 6 & 0.14 & 0 & 0.27($^8_9$) & 6.6($^4_3$) &    & 2.1(4) & \\
L3 & 6 & 0.58 & 0 &     & 6.5($^4_3$) & 1.7($^4_5$) & 4.0($^{0.8}_{1.1}$) & 1.4($^{2.2}_{2.0}$)  \\\hline
\multicolumn{9}{c}{$\beta=6.2$, $c_A=ALPHA$}\\\hline
2 & 6 & 0.14 & 0 & 0.28($^8_9$) & 6.6($^4_3$) &    & 2.2($^3_4$) & \\
L2 & 6 & 0.67 & 0 &     & 6.6($^4_3$) & 1.7($^4_5$) & 4.4($^{0.8}_{1.0}$) &  \\
\end{tabular}
\end{table}

\begin{table}
\caption{$\kappa_c$ determined from $M_{PS}$ and $m_{PCAC}^{imp}$. The 
parameters ``fit'' and $n_k$ are the same as in
table~\protect\ref{chirals}. 
\label{kc}}
\begin{tabular}{c|cc|cc}
&\multicolumn{2}{c|}{$M_{PS}^2$} &    \multicolumn{2}{c}{$m_{PCAC}^{imp}$} \\
& &             &   $c_A=NP$   &  $c_A=NP(ALPHA)$\\\hline
\multicolumn{5}{c}{$\beta=5.93$}\\\hline
fit & 2 & L2 & 2 & 1 \\
$n_{\kappa}$ & 4& 7 & 4 & 4 \\
Q &   0.39 & 0.42 & 0.41 & 0.60 \\
$\kappa_c$ & 0.135252($^{27}_{24}$) & 0.135089($^{16}_{15}$) & 0.135126($^{17}_{15}$) & 0.135099($^5_6$)\\\hline
\multicolumn{5}{c}{$\beta=6.0$}\\\hline
fit & 3 & L2 & - & 1 \\
$n_{\kappa}$ & 6 & 6 & - & 6 \\
Q &   0.77 & 0.28 & - & 0.33 \\
$\kappa_c$ & 0.135291($^{23}_{21}$) & 0.135175($^{16}_{19}$) & - & 0.135185($^{10}_9$)\\\hline
\multicolumn{5}{c}{$\beta=6.2$}\\\hline
fit &  1 & L2 & 1 & 1 \\
$n_{\kappa}$ &  5 & 6 & 6 & 6 \\
Q &   0.35 & 0.36 & 0.34 & 0.31\\
$\kappa_c$ & 0.135820(16) & 0.135786($^{24}_{27}$) & 0.135816(4) & 0.135816(4)\\\hline
\end{tabular}
\end{table}

\begin{table}
\caption{Decay constants and renormalised quark mass at $r^2_0M_{PS}^2=3.0$:
 (a) using $c_A$ from this paper, (b) using $c_A$ as determined by the
 ALPHA collaboration. The two errors given are statistical and
 systematic.  In the case of (a) the statistical errors include those
 from $c_A$, while for (b) the statistical errors from $c_A$ are much
 smaller and are not included. The systematic errors are as follows.
 For $r_0m^{\overline{ms}}_{PCAC}(2GeV)$ from $Z_M$ the errors
 correspond to a $1.1\%$ error in $Z_M$. For
 $r_0m^{\overline{ms}}_{PCAC}(2GeV)$ from $Z_A/Z_P$ we assume a
 $1\alpha^2_P(1/a)$ error in $Z_A/Z_P$~(the mass corrections
 $(b_A-b_P)am_q$ are ignored as they are negligible). Similarly for
 $Z_m$ used for $r_0m^{\overline{ms}}_{\kappa_c}(2GeV)$. For
 $r_0f^{ren}$ the systematic error is dominated by the error in $Z_A$
 obtained by the ALPHA
 collaboration~\protect\cite{nonpertza}.\label{chiralf}}
\begin{tabular}{rlll}
 & $6.2$ & 6.0 & 5.93 \\\hline
\multicolumn{4}{c}{$Z_M$}\\\hline
$r_0m^{\overline{ms}}_{PCAC}(2GeV)$ (a) & 0.240(2)(3) & -  & 0.227(6)(2) \\
(b) & 0.238(1)(3)  & 0.212(2)(2) & -\\\hline
\multicolumn{4}{c}{$Z_A/Z_P$}\\\hline
$r_0m^{\overline{ms}}_{PCAC}(2GeV)$ (a) & 0.234(2)(11) & -  & 0.219(6)(15) \\
(b) & 0.232(1)(11)  & 0.207(2)(13) & -\\\hline\hline
$r_0m^{\overline{ms}}_{\kappa_c}(2GeV)$  &  0.239(7)(13) & $0.232(4)(16)$  & 0.229(3)(17)\\\hline\hline
$r_0f^{ren}$ (a) &  0.443(7)(4)  & -  & 0.440(12)(5) \\
 (b) & 0.439(6)(4)  & 0.402(4)(5) & - 
\end{tabular}
\end{table}

\clearpage

\newpage

\newpage

\begin{figure}
\begin{center}
\centerline{
\epsfxsize=8.0truecm\epsffile{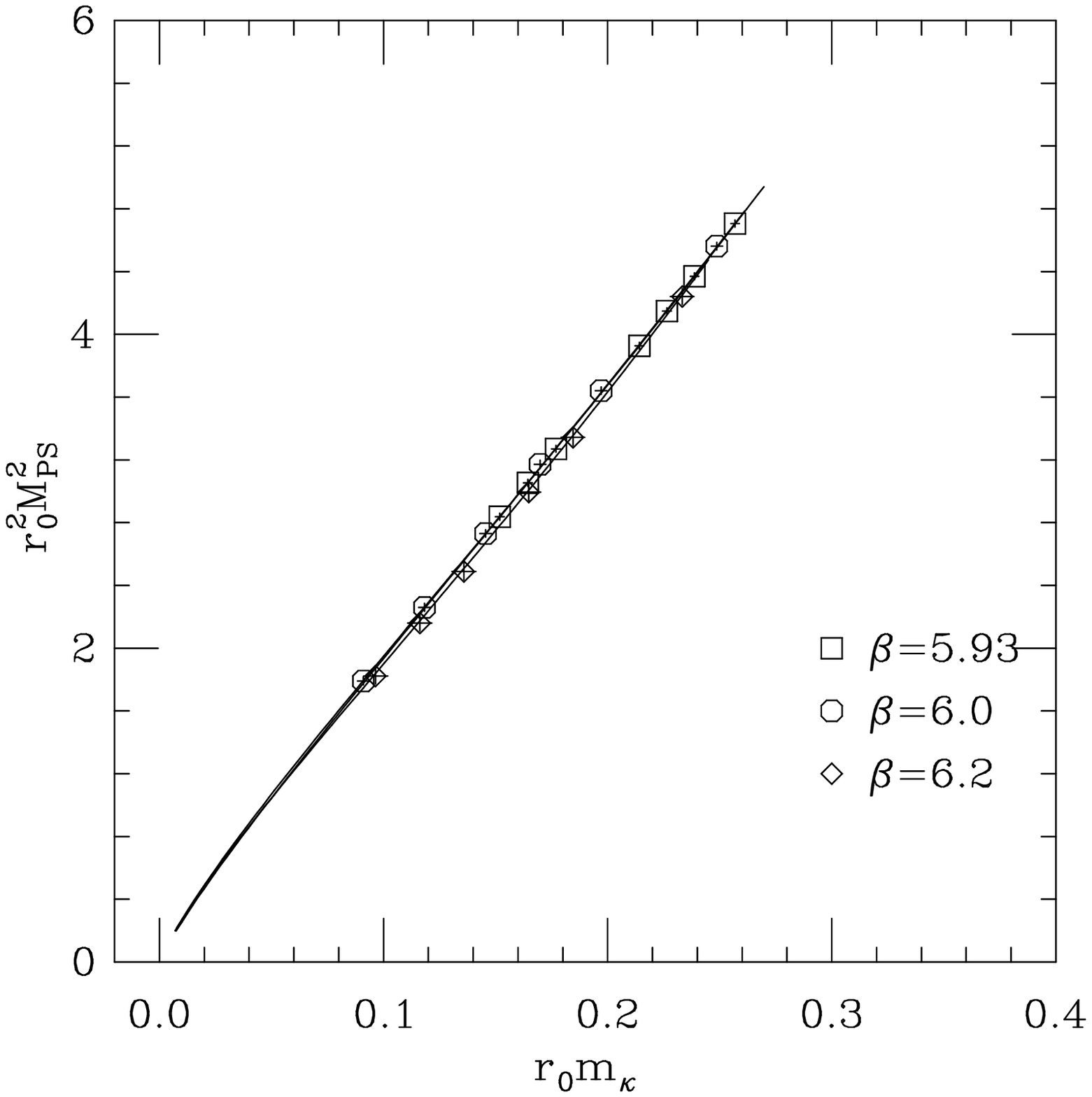}
\epsfxsize=8.0truecm\epsffile{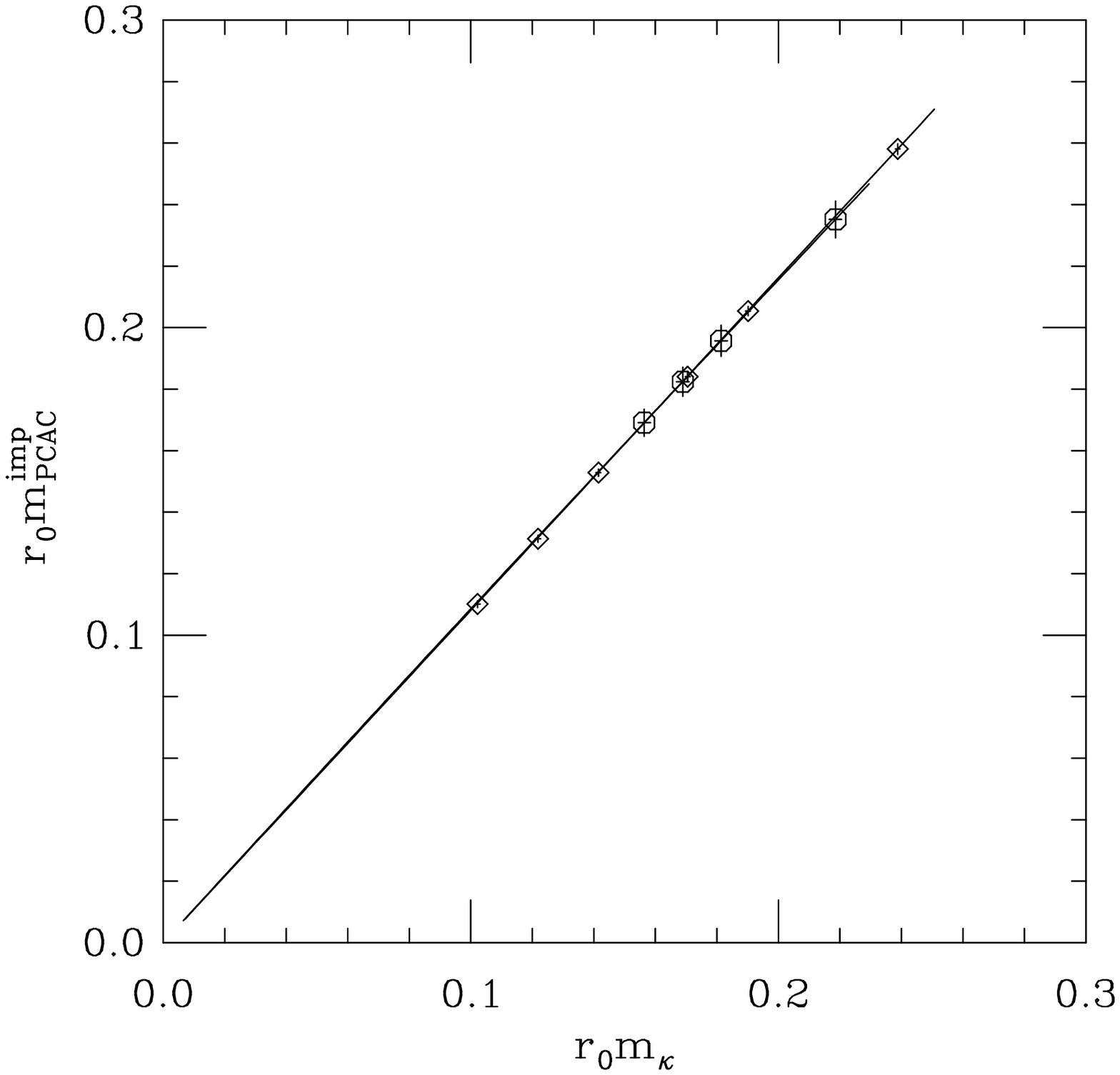}
}
\centerline{(a)\hspace{10cm}(b)}
\centerline{
\epsfxsize=8.0truecm\epsffile{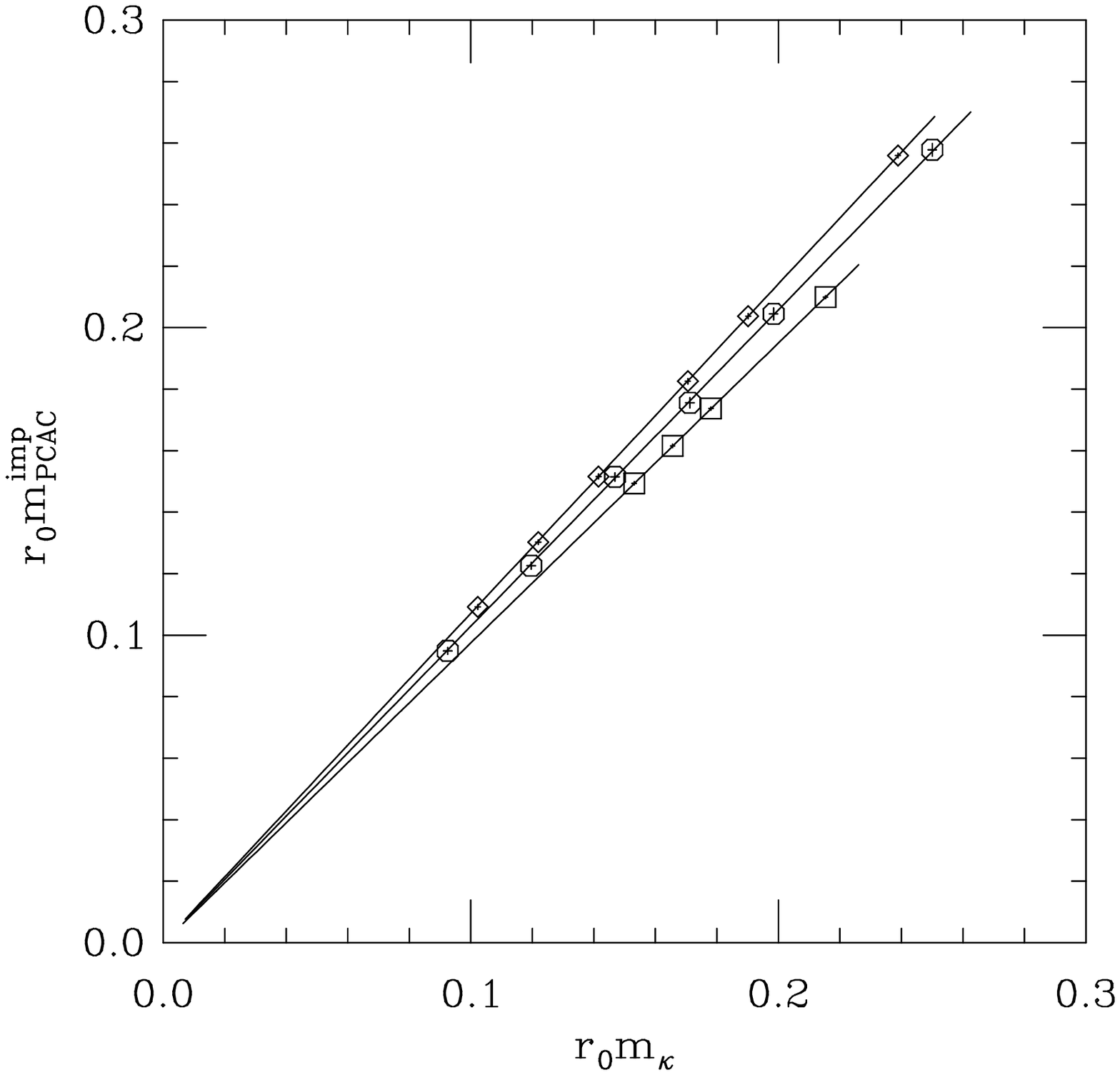}
}
\centerline{(c)}
\caption{ $r_0^2M_{PS}^2$ and $r_0m_{PCAC}^{imp}$ vs $r_0m_\kappa$. $m_\kappa$
is calculated using $\kappa_c$ obtained from the fits shown for each
$\beta$ value. In (b) our results for $c_A$ are used and in (c) the
$c_A$ determined by the ALPHA collaboration.}\label{fig2d}
\end{center}
\end{figure}

\begin{figure}
\begin{center}
\centerline{
\epsfxsize=8.0truecm\epsffile{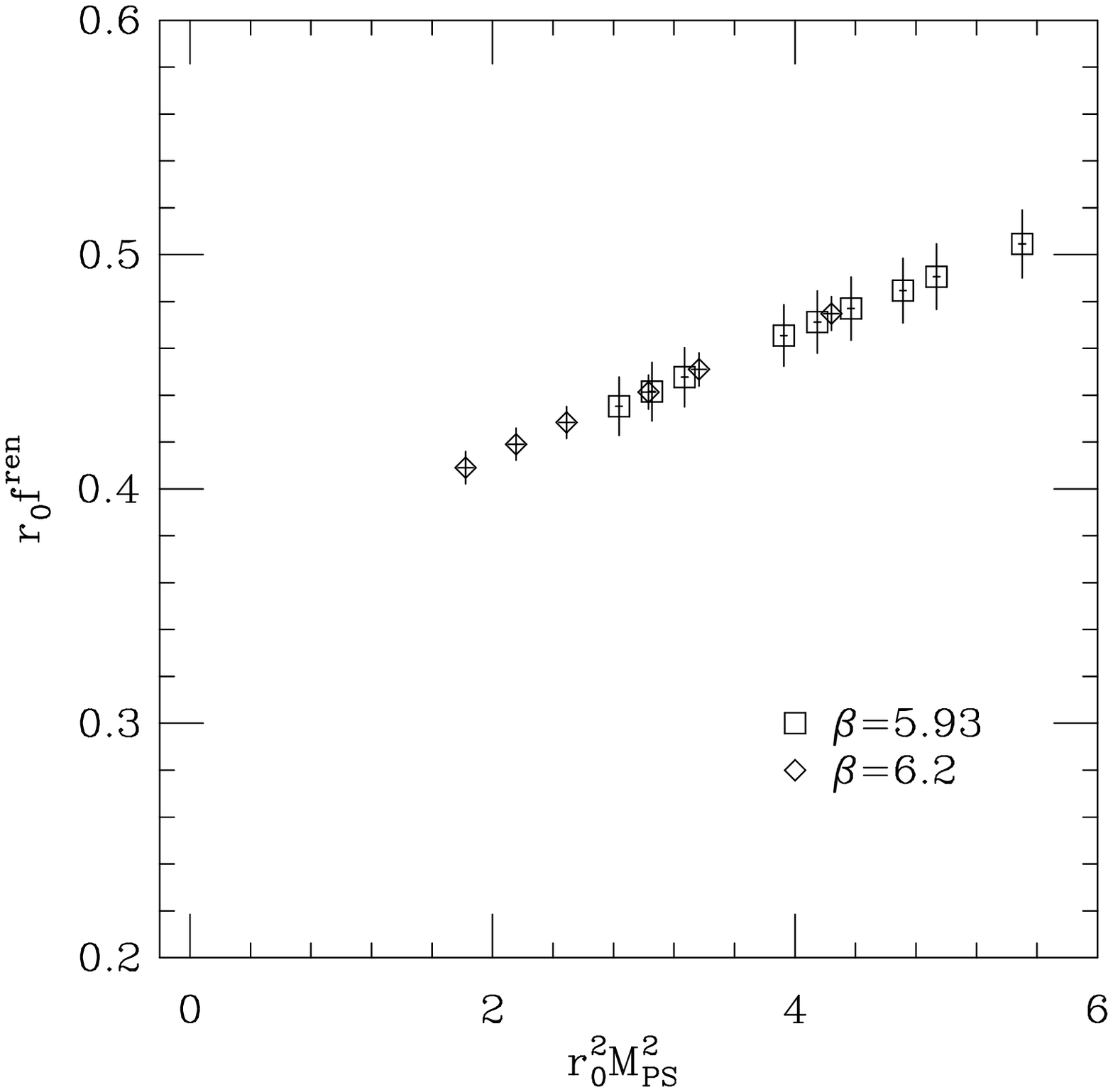}
\epsfxsize=8.0truecm\epsffile{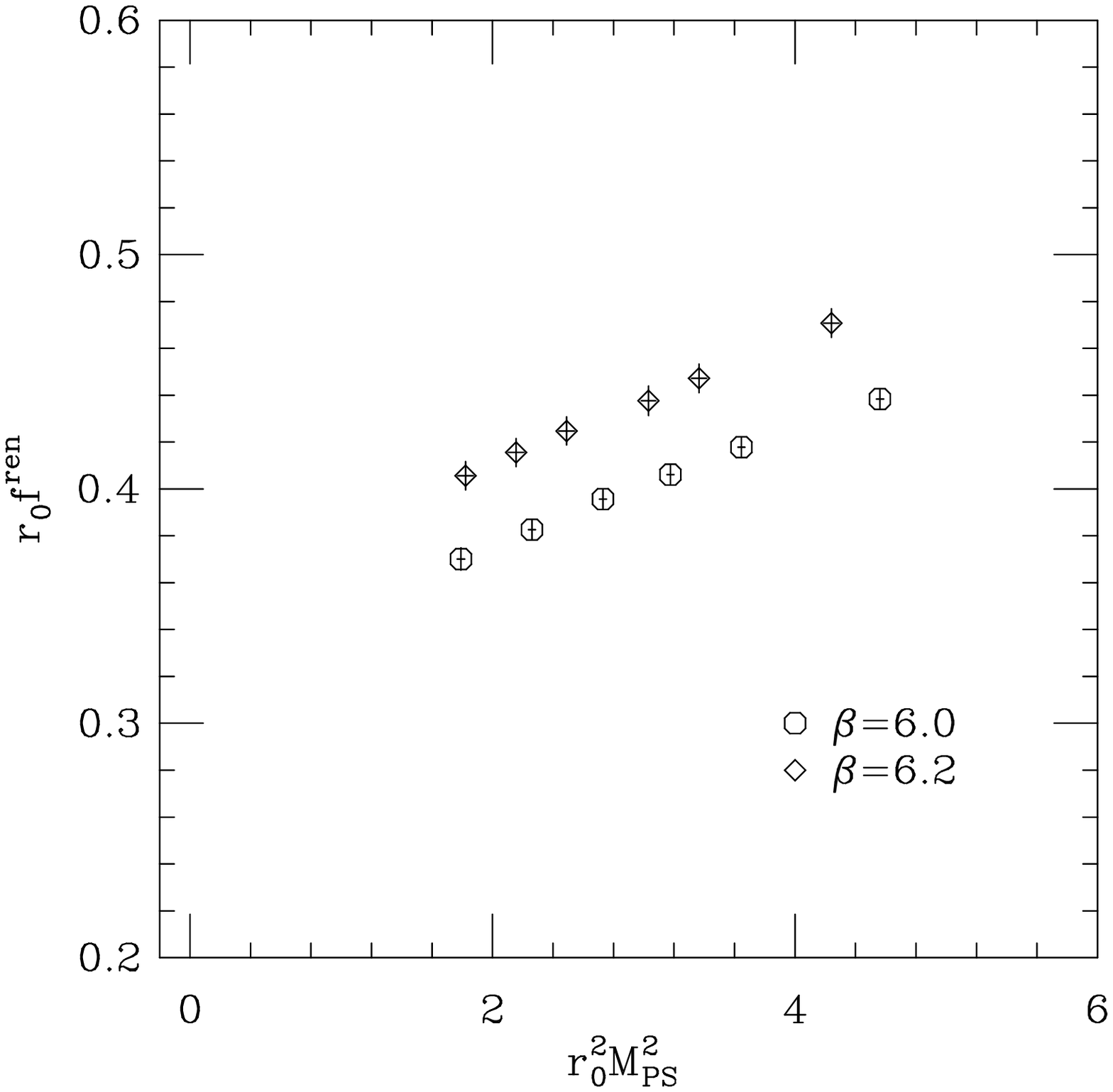}
}
\centerline{(a)\hspace{10cm}(b)}
\centerline{
\epsfxsize=8.0truecm\epsffile{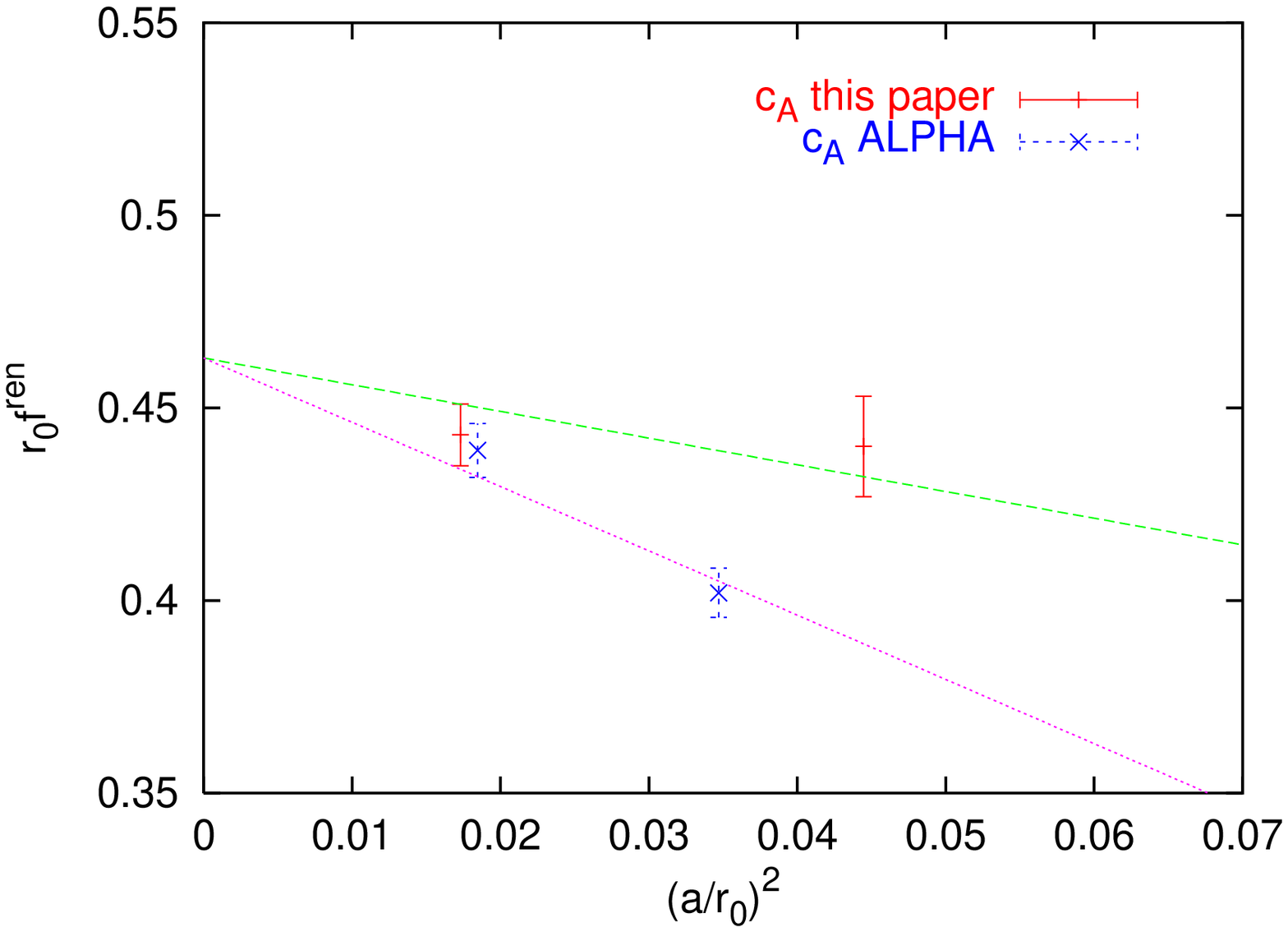}
}
\centerline{(c)}
\caption{ The renormalised pseudoscalar decay constant. (a) $r_0f^{ren}$ vs $r_0^2M_{PS}^2$ using the values of $c_A$ calculated in this paper. The errors are statistical, and include the statistical error of $c_A$. (b) as in (a) using $c_A$ from the ALPHA collaboration. Here the errors on $c_A$ have been neglected. (c) The scaling of $r_0f^{ren}$ for the reference mass $(r_0M_{PS})^2=3.0$. The dashed line indicates a simultaneous fit to the data sets obtained using our $c_A$ values and those from the ALPHA collaboration. The errors include the
uncertainties in the coefficients $b_A$ and $Z_A$. The results at $\beta=6.2$ have been offset slightly for clarity. }\label{fig2e}
\end{center}
\end{figure}

\begin{figure}
\begin{center}
\centerline{
\epsfxsize=8.0truecm\epsffile{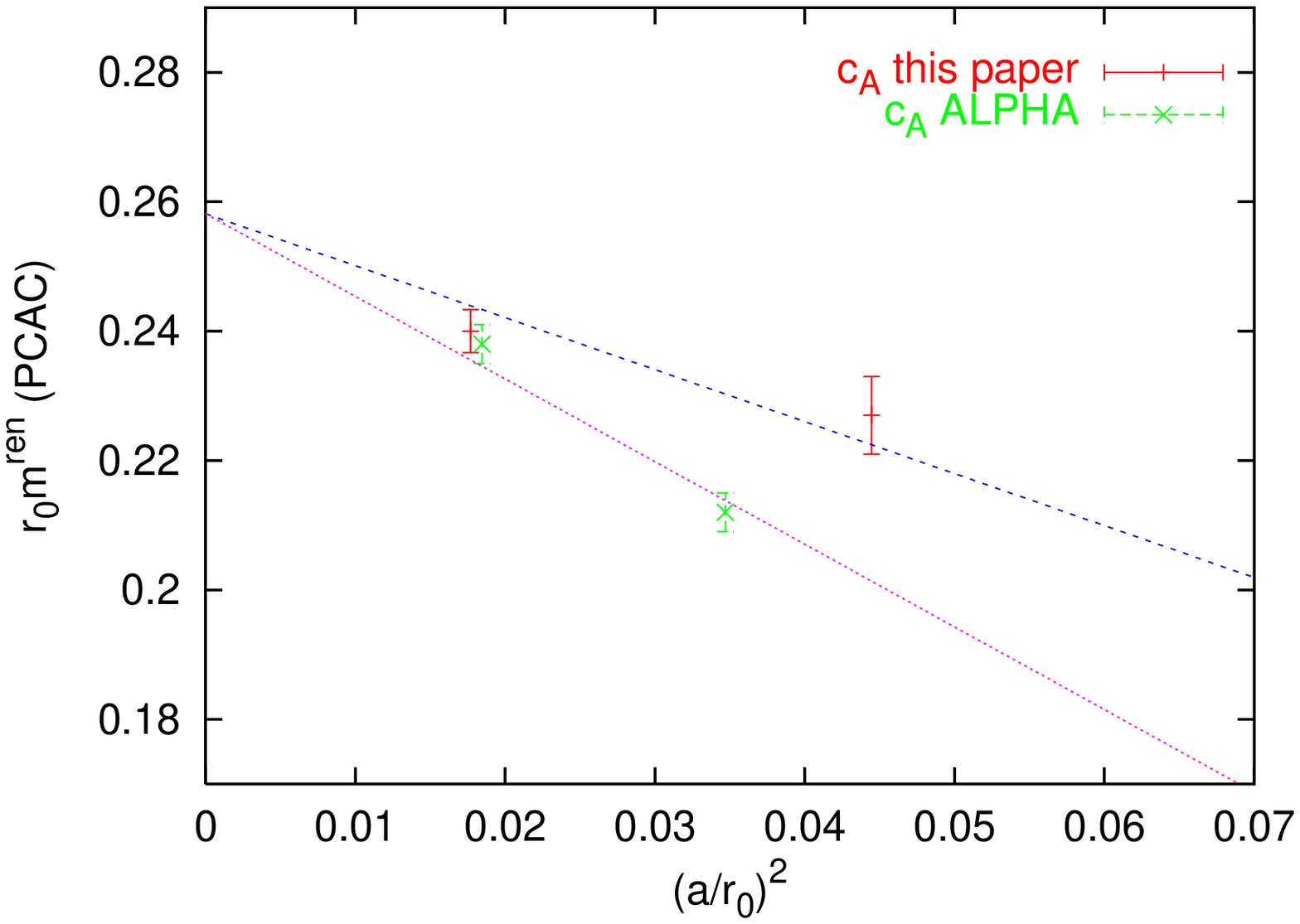}
\epsfxsize=8.0truecm\epsffile{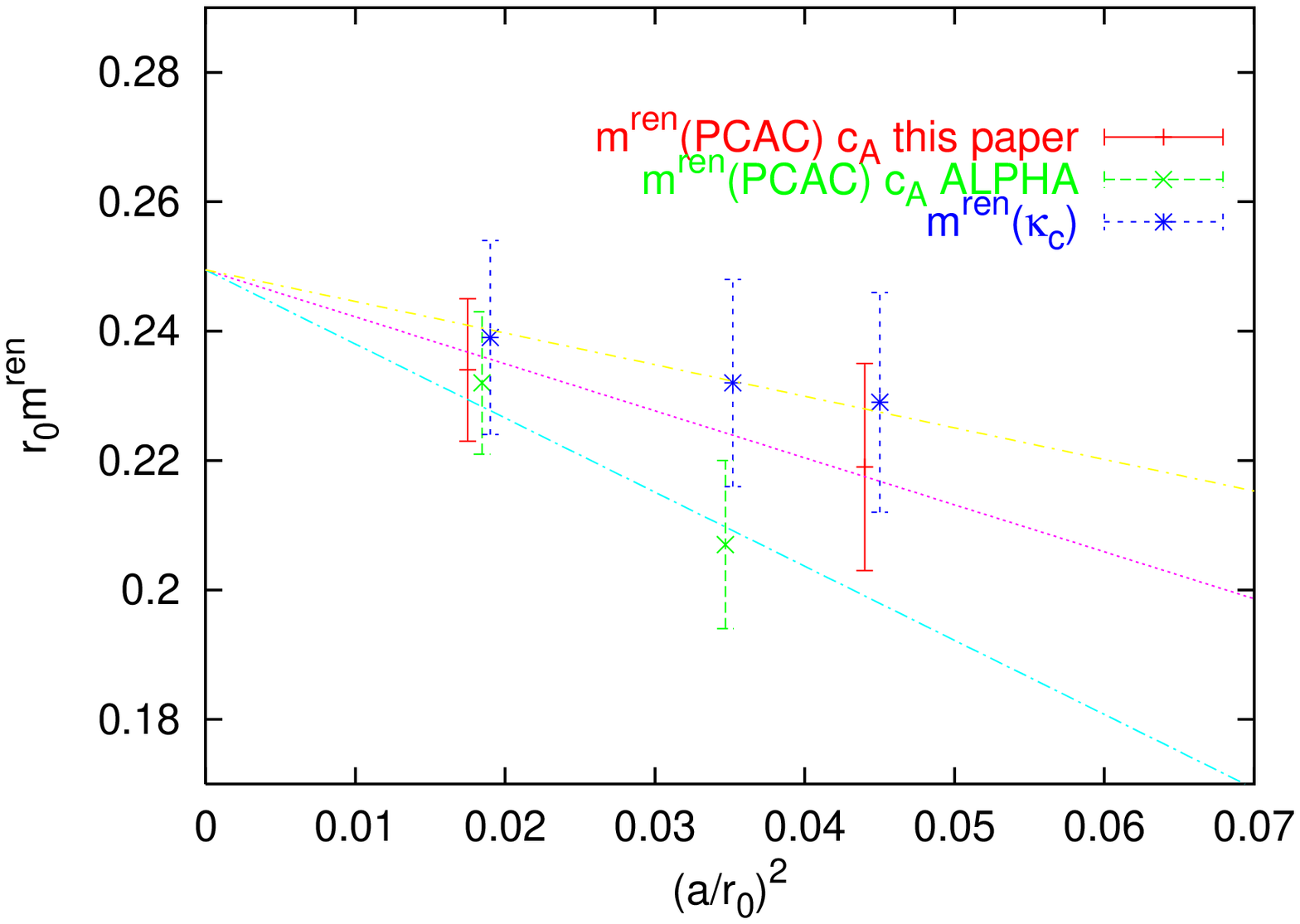}
}
\centerline{(a)\hspace{10cm}(b)}
\caption{ Scaling  of $r_0m^{\overline{MS}}(2GeV)$ for $r^2_0M_{PS}^2=3.0$ (a) obtaining the renormalised quark mass from the bare PCAC quark mass via the renormalisation-group invariant mass~(see equation~\protect\ref{reninv}) and using our values of $c_A$ and those of the ALPHA collaboration. The errors include the uncertainty in $c_A$~(for our $c_A$ values only) and $Z_M$. The dashed line indicates a simultaneous fit to both data sets.  (b) As in (a) but calculating $m^{\overline{MS}}$ directly using the 1-loop perturbative results. This is compared and simultaneously fitted to the 1-loop results for the quark mass extracted using $\kappa_c$, which is independent of $c_A$. The errors include estimates of the perturbative uncertainty in $Z_A/Z_P$. In both plots the data points have been
offset for clarity.}\label{fig2f}
\end{center}
\end{figure}

\end{document}